\documentclass[12pt]{article}
\pdfoutput=1
\usepackage{graphicx,psfrag,epsf,color}
\usepackage[dvipsnames]{xcolor}
\usepackage{amsmath,amssymb,amsfonts, booktabs}
\usepackage{appendix}
\usepackage{multirow}
\usepackage{geometry}
\usepackage{float}
\usepackage{mathrsfs}  
\usepackage{authblk}
\usepackage{array}
\usepackage{cite}
\usepackage{slashed}
\usepackage{graphicx}
\graphicspath{ {Images/} }
\usepackage{rotating}
\usepackage{slashed, cancel, soul}
\usepackage[caption=false]{subfig}
\usepackage[compat=1.1.0]{tikz-feynman} 
\usepackage{hyperref}
\newcounter{MBQ}

\usepackage{tabularx}
\newcolumntype{C}{>{\centering\arraybackslash}X}

\renewcommand{\arraystretch}{1.25}

\def\Eq#1{{Eq.~(\ref{#1})}}
\def\eqs#1#2{{Eqs.~(\ref{#1})--(\ref{#2})}}
\def\fig#1{{Fig.~\ref{#1}}}

\def\Table#1{{Table~\ref{#1}}}
\def\tables#1#2{{Tables~\ref{#1}--\ref{#2}}}
\def\sec#1{{Sect.~\ref{#1}}}

\def\be{\begin{equation}}
\def\ee{\end{equation}}
\def\beq{\begin{eqnarray}}
\def\eeq{\end{eqnarray}}
\newcommand{\bea}{\begin{eqnarray}}
\newcommand{\eea}{\end{eqnarray}}
\newcommand{\beas}{\begin{eqnarray*}}
\newcommand{\eeas}{\end{eqnarray*}}

\newcommand{\nn}{\nonumber}

\newcommand{\bra}[1]{\big\langle{#1}\big\vert}
\newcommand{\ket}[1]{\big\vert{#1}\big\rangle}

\newcommand{\mLam}{m_{\Lambda}}
\newcommand{\mLamB}{m_{\Lambda_b}}

\newcommand{\sLb}{s_{\Lambda_b}}
\newcommand{\sL}{s_\Lambda}

\begin{document}
\allowdisplaybreaks

\begin{titlepage}

\begin{flushright}
{\small
P3H-21-039 \\
SI-HEP-2021-17 \\
Nikhef-2021-012\\
\today \\
}
\end{flushright}

\vskip1cm
\begin{center}
{\bf \boldmath\Large Lepton flavour violation in rare $\Lambda_b$ decays}\\

\end{center}

\vspace{0.5cm}
\begin{center}
{\sc Marzia~Bordone,$^a$ Muslem Rahimi$^b$, K. Keri Vos$^{c,d}$} \\[6mm]
{\it $^a$ Dipartimento di Fisica,
  Universit\`a di Torino \& INFN, Sezione di Torino,\\
I-10125 Torino, Italy \\[0.3cm]
{\it $^b$ Center for Particle Physics Siegen (CPPS), \\
Theoretische Physik 1, Universit{\"a}t Siegen, \\
57068 Siegen, Germany}\\[0.3cm]
{\it $^c$Gravitational 
Waves and Fundamental Physics (GWFP),\\ 
Maastricht University, Duboisdomein 30,\\ 
NL-6229 GT Maastricht, the
Netherlands}\\[0.3cm]

{\it $^d$Nikhef, Science Park 105,\\ 
NL-1098 XG Amsterdam, the Netherlands}}
\end{center}

\vspace{0.6cm}
\begin{abstract}
\vskip0.2cm\noindent
Lepton flavour violation (LFV) naturally occurs in many new physics models, specifically in those explaining the $B$ anomalies. While LFV has already been studied for mesonic decays, it is important to consider also baryonic decays mediated by the same quark transition. In this paper, we study LFV in the baryonic $\Lambda_b \to \Lambda \ell_1 \ell_2$ using for the first time a full basis of New Physics operators. We present expected bounds on the branching ratio in a model-independent framework and using two specific new physics models. Finally, we point out the interplay and orthogonality between the baryonic and mesonic LFV searches. 
\end{abstract}
\end{titlepage}



\section{Introduction}

The incredible joint theoretical and experimental effort carried out in the last years allows us to probe the Standard Model (SM) of particle physics with an unprecedented precision. This brought to light some deviations between theoretical predictions and experimental measurements in semileptonic $B$ meson decays \cite{Aaij:2021vac,Aaij:2020ruw,Aaij:2020nrf,Aaij:2014ora,Aaij:2017vbb,Aaij:2019wad,Aaij:2015oid,Lees:2012xj,Lees:2013uzd,Aaij:2015yra,Hirose:2016wfn,Hirose:2017dxl,Aaij:2017uff,Aaij:2017deq,Abdesselam:2019dgh}. These discrepancies, the so-called $B$ anomalies, hint at Lepton Flavour Universality (LFU) violation. This is quite surprising, as LFU is one of the foundation of the SM. 

The $B$ anomalies can be split into two classes: \textit{i}) deviations in $\mu/e$ universality in $b\to s\ell^+\ell^-$ and \textit{ii}) deviations in $\tau$ vs. light leptons universality in the $b\to c\ell\bar\nu$ transitions. These exciting findings may indicate the presence of New Physics (NP) particles, and have inspired a plethora of theoretical and experimental work. The NP explanations for $B$ anomalies span a broad class of new heavy particles, from vectors to scalars states\cite{Alonso:2015sja,Calibbi:2015kma,DiLuzio:2017vat,Calibbi:2017qbu,Barbieri:2017tuq,Blanke:2018sro,DiLuzio:2018zxy,Faber:2018qon,Heeck:2018ntp,Angelescu:2018tyl,Schmaltz:2018nls,Greljo:2018tzh,Fornal:2018dqn,Baker:2019sli,Cornella:2019hct,DaRold:2019fiw,Bordone:2017bld,Bordone:2018nbg,Bordone:2019uzc,Marzocca:2018wcf,Becirevic:2018afm,Bigaran:2019bqv,Crivellin:2019dwb,Saad:2020ihm,Gherardi:2020qhc,Babu:2020hun,Crivellin:2017zlb,Buttazzo:2017ixm,Bordone:2020lnb,Cornella:2021sby,Marzocca:2021azj,Greljo:2021xmg,Davighi:2021oel,Alvarado:2021nxy,FileviezPerez:2021xfw,Cen:2021iwv,Chen:2021vzk,Nomura:2021oeu,Lee:2021jdr,Arcadi:2021cwg,Barbieri:2021wrc,Hiller:2021pul,Angelescu:2021lln,Alda:2020okk,Arcadi:2021glq}. However, a common feature in all of these models is the prediction of sizeable effects for Lepton Flavour Violating (LFV) $B$, $\tau$ and $\mu$ decays. Current upper bounds on these modes largely constrain the allowed parameter space for NP models, and upcoming experimental analyses will be fundamental to corroborate or falsify these NP hypotheses.

When searching for LFV decays mediated by $b\to s\ell_1\ell_2$ transitions, it is crucial to consider both mesonic and baryonic decays. Although mediated by the same underlying partonic transition, these two types of decays provide orthogonal information on possible NP models. A striking example of this is the NP analysis of $\Lambda_b\to\Lambda\mu^+\mu^-$ decays \cite{Blake:2019guk}, which shows that even though $B\to K\mu^+\mu^-$ angular distribution seems to be affected by some short-distance NP \cite{Alguero:2021anc,Altmannshofer:2021qrr,Hurth:2021nsi,Ciuchini:2019usw}, the latter is not visible in the $\Lambda_b\to\Lambda\mu^+\mu^-$ angular observables.  It is therefore natural to assume that $\Lambda_b\to\Lambda \ell_1^-\ell_2^+$ decays provide complementary information compared to their mesonic counterparts $B^+\to K^+\ell_1\ell_2$ or $\bar B_s\to\ell_1\ell_2$ decays. More precisely, the spin structure of the $\Lambda_b\to \Lambda$ decays induces a richer set of hadronic matrix elements. Therefore, $\Lambda_b\to \Lambda$ decays probe different parameter space than their mesonic counter parts. In addition, the $\Lambda_b$ baryon is copiously produced at LHCb and the dataset collected with Run 1 and Run 2 allow to make precision measurement of observables constructed from $\Lambda_b$ decays (see e.g. \cite{Bediaga:2018lhg}). 

In this paper, we calculate for the first time the angular distribution of $\Lambda_b\to\Lambda \ell_1^-\ell_2^+$ decays using a full base of NP operators (partial results are available in \cite{Sahoo:2016nvx,Das:2019omf}). To achieve this, we use the decomposition of the $\Lambda_b\to\Lambda$ hadronic matrix elements in \cite{Feldmann:2011xf} and the lattice QCD determination of the corresponding form factors \cite{Detmold:2016pkz}. We then use model independent constraints to derive upper bounds for the branching ratio of $\Lambda_b\to\Lambda \ell_1^-\ell_2^+$ decays and specific models to provide predictions in a few scenarios \cite{Cornella:2021sby,Bordone:2020lnb}.\\
This paper is organised as follows: in \sec{sec:2} we highlight the main steps of our calculation and provide numerical results in a generic scenario. In \sec{sec:3} we use constraints on various LFV mesonic decays to put bounds on the branching ratio of $\Lambda_b\to\Lambda \ell_1^-\ell_2^+$, for different choices of leptons in the final state and make predictions for specific models. We conclude in \sec{sec:4}.
\\

\section{The angular distribution of \boldmath$ \Lambda_b\to\Lambda \, \ell_1^{-} \ell_2^{+}$ }
\label{sec:2}
In this Section, we introduce the concepts that we need for the study of phenomenological aspects in \sec{sec:3}. 

We consider the following effective Hamiltonian for LFV $b\to s\ell_1^-\ell_2^+$ transitions:
\begin{equation}
\label{eq:heff}
\mathcal{H}_{\mathrm{eff}}=-\frac{4 G_F}{\sqrt{2}}V_{tb}V_{ts}^* \frac{\alpha_\mathrm{em}}{4\pi} \sum_{i=9,10,S,P} \Bigg{(}C^{\ell_1\ell_2}_i(\mu)\mathcal{O}^{\ell_1\ell_2}_i(\mu)+C_i^{\prime\ell_1\ell_2}(\mu)\mathcal{O}_i^{\prime\ell_1\ell_2}(\mu)\Bigg{)},
\end{equation}
where the relevant operators are defined by
\begin{align}
\label{eq:ops}
\mathcal{O}_{9}^{\ell_1\ell_2} &=(\bar{s}\gamma_\mu P_L b)(\bar{\ell_1}\gamma^\mu\ell_{2}),& \mathcal{O}_{10}^{\ell_1\ell_2}& = (\bar{s}\gamma_\mu P_L b)(\bar{\ell_1}\gamma^\mu\gamma^5\ell_{2}),\nn \\
\mathcal{O}_{S}^{\ell_1\ell_2} &= (\bar{s}P_R b)(\bar{\ell_1}\ell_{2}),& \mathcal{O}_{P}^{\ell_1\ell_2} &= (\bar{s}P_R b)(\bar{\ell_1}\gamma_5\ell_{2})\,, \\  \mathcal{O}_{T}^{\ell_1\ell_2} &= (\bar{s} \sigma^{\mu \nu} b) (\bar{\ell}_{1} \sigma_{\mu \nu} \ell_2)\,,&  \mathcal{O}_{T5}^{\ell_1\ell_2} &= (\bar{s} \sigma^{\mu \nu} b) (\bar{\ell}_{1} \sigma_{\mu \nu}\gamma_5 \ell_2)\,,\nonumber
\end{align}
and the operators with flipped chirality $\mathcal{O}^{\prime\ell_1\ell_2}_{i}$ are obtained from  $\mathcal{O}^{\ell_1\ell_2}_{i}$ by replacing $P_L \leftrightarrow P_R$, where $P_{L/R}=\frac{1}{2}(1\mp \gamma_5)$.   
Notice that the operator $\mathcal{O}_7$:
\begin{equation}
    \mathcal{O}_{7} = \frac{m_b}{e}(\bar{s}\sigma_{\mu\nu}P_R b)F^{\mu\nu} 
\end{equation}
cannot generate LFV contributions due to the universality of electromagnetic interactions. 
We parametrise the hadronic matrix elements for $\Lambda_b(p,\sLb)\to\Lambda(k,\sL)$ decays using an helicity decomposition \cite{Feldmann:2011xf,Detmold:2016pkz,Datta:2017aue,Boer:2014kda}:
\begin{align}
 \nonumber \langle \Lambda(k,\sL) | \bar{s} \,\gamma^\mu\, b | \Lambda_b(p,\sLb) \rangle =\,&+
 \bar{u}_\Lambda(k,\sL) \bigg[ f_0(q^2)\: (m_{\Lambda_b}-m_\Lambda)\frac{q^\mu}{q^2} \\
 \nonumber & + f_+(q^2) \frac{m_{\Lambda_b}+m_\Lambda}{s_+}\left( p^\mu + k^{\mu} - (m_{\Lambda_b}^2-m_\Lambda^2)\frac{q^\mu}{q^2}  \right) \\
 &+ f_\perp(q^2) \left(\gamma^\mu - \frac{2m_\Lambda}{s_+} p^\mu - \frac{2 m_{\Lambda_b}}{s_+} k^{ \mu} \right) \bigg] u_{\Lambda_b}(p,\sLb), \label{eq:HMEL1}\\
 \nonumber \langle \Lambda(k,\sL) | \bar{s} \,\gamma^\mu\gamma_5\, b | \Lambda_b(p,\sLb) \rangle =\,&
 -\bar{u}_\Lambda(k,\sL) \:\gamma_5 \bigg[ g_0(q^2)\: (m_{\Lambda_b}+m_\Lambda)\frac{q^\mu}{q^2} \\
 \nonumber & + g_+(q^2)\frac{m_{\Lambda_b}-m_\Lambda}{s_-}\left( p^\mu + k^{\mu} - (m_{\Lambda_b}^2-m_\Lambda^2)\frac{q^\mu}{q^2}  \right) \nonumber\\
 & + g_\perp(q^2) \left(\gamma^\mu + \frac{2m_\Lambda}{s_-} p^\mu - \frac{2 m_{\Lambda_b}}{s_-} k^{\mu} \right) \bigg]  u_{\Lambda_b}(p,\sLb)\,, \label{eq:HMEL2}\\
\bra{\Lambda(k,s_\Lambda)}\bar{s} i \sigma^{\mu\nu} b\ket{\Lambda_b(p,s_{\Lambda_b})}= \, & + \bar{u}_{\Lambda}(k, s_\Lambda) \Big\{2h_+(q^2)\frac{p^\mu k^{ \nu}-p^\nu k^{\mu}}{s_{+}} \nonumber\\
&+h_\perp (q^2)\Big[\frac{m_{\Lambda_{b}}+m_{\Lambda}}{q^2}(q^\mu \gamma^\nu -q^\nu \gamma^\mu)-2\left(\frac{1}{q^2}+\frac{1}{s_{+}}\right)(p^\mu k^{\nu}-p^\nu k^{\mu}) \Big] \nonumber\\
&+\widetilde{h}_+ (q^2)\Big[i\sigma^{\mu \nu}-\frac{2}{s_{-}}(m_{\Lambda_{b}}(k^{\mu}\gamma^\nu -k^{\nu}\gamma^\mu)\nonumber\\
&-m_{\Lambda}(p^\mu \gamma^\nu -p^\nu \gamma^\mu)+p^\mu k^{\nu}-p^\nu k^{\mu}) \Big] \nonumber\\
&+\widetilde{h}_\perp(q^2) \frac{m_{\Lambda_{b}}-m_{\Lambda}}{q^2 s_{-}}\Big[(m_{\Lambda_{b}}^2-m_{\Lambda}^2-q^2)(\gamma^\mu p^\nu - \gamma^\nu p^\mu)\nonumber\\
& -(m_{\Lambda_{b}}^2-m_{\Lambda}^2+q^2)(\gamma^\mu k^{\nu}-\gamma^\nu k^{\mu})\nonumber\\
&+2(m_{\Lambda_{b}}-m_{\Lambda})(p^\mu k^{\nu}-p^\nu k^{\mu}) \Big]
\Big\} u_{\Lambda_b}(p, s_{\Lambda_b})\, \label{eq:HMEL3}
\end{align}
with $q=p-k$, $s_\pm =(m_{\Lambda_b} \pm m_\Lambda)^2-q^2$, and $\sLb$ and $\sL$ are the spin of the $\Lambda_b$ and $\Lambda$ baryons, respectively. Applying equations of motion to  \eqs{eq:HMEL1}{eq:HMEL3}, we obtain the following matrix elements for scalar and pseudoscalar operators:
\begin{align}
    \langle \Lambda(k,\sL)| \bar{s}    \, b|\Lambda_b(p,\sLb)\rangle =\,& \frac{\mLamB-\mLam}{m_b(\mu)-m_s(\mu)}f_0\bar{u}_\Lambda(k,\sL) u_{\Lambda_b}(p,\sLb)\,, \\
     \langle \Lambda(k,\sL)| \bar{s}  \gamma_5   b|\Lambda_b(p,\sLb)\rangle =\,& \frac{\mLamB+\mLam}{m_b(\mu)+m_s(\mu)}g_0\bar{u}_\Lambda(k,\sL)\gamma_5 u_{\Lambda_b}(p,\sLb)\,,
\end{align}
which agree with the expressions in Ref.~\cite{Feldmann:2011xf}. In the following, we take the masses in $\overline{\rm MS}$ using $\overline{m}_b(\overline{m}_b)=4180$ MeV \cite{Zyla:2020zbs} and $\overline{m}_s(\overline{m}_b)=78\,\mathrm{MeV}$\, \cite{Chetyrkin:2000yt}.

\subsection{Differential decay width and numerical analysis}
We decompose the spin-independent double-differential decay width as
\begin{equation}
    \frac{1}{\Gamma^{(0)}}\frac{d\Gamma(\Lambda_b(p,\sLb)\to\Lambda(k,\sL) \ell_1^-(p_1) \ell_2^+(p_2))}{d\cos\theta d q^2}=  a + b \cos\theta_\ell + c \cos^2\theta_\ell\,,
\end{equation}
with $\Gamma^{(0)}=\frac{ \alpha_\mathrm{em}^2 G_\mathrm{F}^2 |V_{tb}V_{ts}^*|^2}{2048\pi^5\mLamB^3 q^2}\sqrt{\lambda_H}\sqrt{\lambda_L}$. We define $\lambda_H\equiv\lambda(\mLamB^2, \mLam^2, q^2)$ and $\lambda_L\equiv\lambda(q^2, m_{\ell_1}^2, m_{\ell_2}^2)$, where $\lambda$ is the usual K\"all\'en function defined as $\lambda(a,b,c) = a^2+b^2+c^2-2 a (b+c)-2bc$. Here $\cos\theta_\ell$ is the helicity angle in the dilepton frame as defined in Appendix~\ref{app:detkin}. The coefficients $a$, $b$ and $c$ are one of the main result of this work and have been calculated using the operator base in \Eq{eq:heff} and the decomposition for the hadronic matrix elements in \eqs{eq:HMEL1}{eq:HMEL3}. We find: 
\begin{align}
    a =&-\frac{1}{q^2} \bigg\{|f_0|^2\frac{(\mLamB-\mLam)^2}{q^2}s_+[|C^{\ell_1\ell_2}_{10+}|^2(m_{\ell_{1}}+m_{\ell_{2}})^2q_{- } + |C^{\ell_1\ell_2}_{9+}|^2 (m_{\ell_{1}}-  m_{\ell_{2}})^2q_{+ }] \nonumber\\
    &+|f_\perp|^2 s_-[|C^{\ell_1\ell_2}_{10+}|^2 (\lambda_L+2q^2 q_{+ })+|C^{\ell_1\ell_2}_{9+}|^2(\lambda_L+2q^2 q_{-})]\nonumber\\
    & + |f_+|^2(\mLamB+\mLam)^2 s_-[|C^{\ell_1\ell_2}_{10+}|^2 \,q_{+} +|C^{\ell_1\ell_2}_{9+}|^2 \, q_{-}]\nonumber\\
    &+|g_0|^2\frac{(\mLamB+\mLam)^2}{q^2}s_-[|C^{\ell_1\ell_2}_{10-}|^2(m_{\ell_{1}} +  m_{\ell_{2}})^2q_{- } +|C^{\ell_1\ell_2}_{9-}|^2(m_{\ell_{1}}-  m_{\ell_{2}})^2q_{+ }] \nonumber\\
    &+|g_\perp|^2 s_+[|C^{\ell_1\ell_2}_{10-}|^2(\lambda_L+2q^2 q_{+ })+|C^{\ell_1\ell_2}_{9-}|^2(\lambda_L+2q^2 q_{- })]\nonumber\\
    &+|g_+|^2(\mLamB-\mLam)^2 s_+[|C^{\ell_1\ell_2}_{10-}|^2 q_{+ } +|C^{\ell_1\ell_2}_{9-}|^2 q_{- }]\nonumber\\
    &+16|h_+|^2s_-[(m_{\ell_1}+m_{\ell_2})^2 q_- |C^{\ell_1\ell_2}_{T}|^2+(m_{\ell_1}-m_{\ell_2})^2 q_+|C^{\ell_1\ell_2}_{T5}|^2] \nonumber\\
    &+16|\tilde h_+|^2s_+[(m_{\ell_1}-m_{\ell_2})^2 q_- |C^{\ell_1\ell_2}_{T}|^2+(m_{\ell_1}+m_{\ell_2})^2 q_+ |C^{\ell_1\ell_2}_{T5}|^2]\nonumber \\ &+16|h_\perp|^2\frac{s_-}{q^2}(\mLamB+\mLam)^2[q_-((m_{\ell_1}+m_{\ell_2})^2+q^2)|C^{\ell_1\ell_2}_{T}|^2+q_+((m_{\ell_1}+m_{\ell_2})^2+q^2)|C^{\ell_1\ell_2}_{T5}|^2] \nonumber \\
    &+16|\tilde h_\perp|^2\frac{s_+}{q^2}(\mLamB-\mLam)^2[q_+((m_{\ell_1}-m_{\ell_2})^2+q^2)|C^{\ell_1\ell_2}_{T}|^2+q_-((m_{\ell_1}+m_{\ell_2})^2+q^2)|C^{\ell_1\ell_2}_{T5}|^2]\bigg\}\nonumber\\
    &-(q_- |C_{P-}^{\ell_1\ell_2}|^2+q_+|C_{S-}^{\ell_1\ell_2}|^2) \frac{s_- (\mLamB+\mLam)^2}{(m_b+m_s)^2}|g_0|^2
    -(q_- |C_{P+}^{\ell_1\ell_2}|^2+q_+|C_{S+}^{\ell_1\ell_2}|^2)\frac{s_+ (\mLamB-\mLam)^2}{(m_b-m_s)^2}|f_0|^2\nonumber \\
    &-\frac{2s_+}{q^2} \frac{(\mLamB-\mLam)^2}{m_b-m_s}[\mathrm{Re}(C^{\ell_1\ell_2}_{10+} C^{*\ell_1\ell_2}_{P+})(m_{\ell_{1}}+m_{\ell_{2}})q_-+\mathrm{Re}(C^{\ell_1\ell_2}_{9+} C^{*\ell_1\ell_2}_{S+})(m_{\ell_{1}}-m_{\ell_{2}})q_+]|f_0|^2 \nonumber\\
    &-\frac{2s_-}{q^2} \frac{(\mLamB+\mLam)^2}{m_b+m_s}[\mathrm{Re}(C^{\ell_1\ell_2}_{10-} C^{*\ell_1\ell_2}_{P-})(m_{\ell_{1}}+m_{\ell_{2}})q_-+\mathrm{Re}(C^{\ell_1\ell_2}_{9-} C^{*\ell_1\ell_2}_{S-})(m_{\ell_{1}}-m_{\ell_{2}})q_+]|g_0|^2 \nonumber \\
    & - \frac{8}{q^2} \bigg\{ (m_{\ell_1} +m_{\ell_2})(\mLamB +\mLam) s_- q_- \text{Re}(C_{9+}^{\ell_{1} \ell_2} \,C_T^{* \ell_{1} \ell_2}) \big[ \text{Re}(f_+ \,h_+^{*}) +2\text{Re}(f_\perp \,h_\perp^{*}) \big] \nonumber \\
    & +  (m_{\ell_1} - m_{\ell_2}) (\mLamB -\mLam) s_+ q_+ \text{Re}(C_{10-}^{\ell_{1} \ell_2} \,C_T^{* \ell_{1} \ell_2}) \big[ \text{Re}(g_+ \, \tilde{h}_+^{*}) +2\text{Re}(g_\perp \,\tilde{h}_\perp^{*}) \big] \nonumber \\
    & + (m_{\ell_1} +m_{\ell_2}) (\mLamB - \mLam) s_+ q_- \text{Re}(C_{9-}^{\ell_{1} \ell_2} \,C_{T5}^{* \ell_{1} \ell_2}) \big[ \text{Re}(g_+ \,\tilde{h}_+^{*} +2\text{Re}(g_\perp \, \tilde{h}_\perp^{*})) \big] \nonumber \\
    & +(m_{\ell_1} - m_{\ell_2}) (\mLamB + \mLam) s_- q_+ \text{Re}(C_{10+}^{\ell_{1} \ell_2} \,C_{T5}^{* \ell_{1} \ell_2}) \big[ \text{Re}(f_+ \,\tilde{h}_+^{*} +2\text{Re}(f_\perp \, \tilde{h}_\perp^{*})) \big] \bigg\}\,, \label{eq:a}\\
    b =& \frac{2}{q^4} \big[\mathrm{Re}  (f_0 f_+^*) (|C^{\ell_1\ell_2}_{9+}|^2+|C^{\ell_1\ell_2}_{10+}|^2) +\mathrm{Re}  (g_0 g_+^*) (|C^{\ell_1\ell_2}_{9-}|^2+|C^{\ell_1\ell_2}_{10-}|^2) \big]\sqrt{\lambda_H\lambda_L}(m_{\ell_{2}}^2-m_{\ell_{1}}^2)(\mLamB^2-\mLam^2)\nonumber\\
    &-4[ \mathrm{Re}(C^{\ell_1\ell_2}_{9-} C^{*\ell_1\ell_2}_{10+})+\mathrm{Re}(C^{\ell_1\ell_2}_{9+}C^{*\ell_1\ell_2}_{10-})]\mathrm{Re}  (f_\perp g_\perp^*)\sqrt{\lambda_H\lambda_L}\nonumber\\
    &-\frac{2(\mLamB^2-\mLam^2)}{q^2(m_b-m_s)}\sqrt{\lambda_H\lambda_L}[\mathrm{Re}(C^{\ell_1\ell_2}_{10+} C^{*\ell_1\ell_2}_{P+})(m_{\ell_{1}}-m_{\ell_{2}})+\mathrm{Re}(C^{\ell_1\ell_2}_{9+} C^{*\ell_1\ell_2}_{S+})(m_{\ell_{1}}+m_{\ell_{2}})]\mathrm{Re}(f_0 f_+^*) \nonumber\\
     &-\frac{2(\mLamB^2-\mLam^2)}{q^2(m_b+m_s)}\sqrt{\lambda_H\lambda_L}[\mathrm{Re}(C^{\ell_1\ell_2}_{10-} C^{*\ell_1\ell_2}_{P-})(m_{\ell_{1}}-m_{\ell_{2}})+\mathrm{Re}(C^{\ell_1\ell_2}_{9-} C^{*\ell_1\ell_2}_{S-})(m_{\ell_{1}}+m_{\ell_{2}})]\mathrm{Re}(g_0 g_+^*) \nonumber\\
     &+\frac{64}{q^4}(|C^{\ell_1\ell_2}_{T}|^2+|C^{\ell_1\ell_2}_{T5}|^2)(m_{\ell_2}^2-m_{\ell_1}^2)(\mLamB^2-\mLam^2)\sqrt{\lambda_H \lambda_L}\text{Re}(h_\perp \tilde h_\perp^*) \nonumber \\
     & - \frac{8}{q^2} \bigg \{ \text{Re}(C_{9+}^{\ell_1 \ell_2}\, C_T^{*\ell_1 \ell_2}) (m_{\ell_1} -m_{\ell_2}) (\mLamB - \mLam) \sqrt{\lambda_H \lambda_L} \big[ \text{Re}(f_0 h_+^{*}) + 2\text{Re}( f_\perp \tilde{h}_\perp^{*}) \big] \nonumber \\
     &+ \text{Re}(C_{10-}^{\ell_1 \ell_2}\, C_T^{*\ell_1 \ell_2}) (m_{\ell_1} +m_{\ell_2}) (\mLamB + \mLam) \sqrt{\lambda_H \lambda_L} \big[ \text{Re}(g_0 \tilde{h}_+^{*}) + 2\text{Re}( g_\perp h_\perp^{*}) \big] \nonumber \\
     & + q^2 \frac{(\mLamB - \mLam)}{(m_b- m_s)} \text{Re}(C_{S+}^{\ell_1 \ell_2}\, C_T^{*\ell_1 \ell_2}) \sqrt{\lambda_H \lambda_L} \text{Re}(f_0 \,h_+^{*}) +q^2 \frac{(\mLamB + \mLam)}{(m_b + m_s)} \text{Re}(C_{P-}^{\ell_1 \ell_2}\, C_T^{*\ell_1 \ell_2}) \sqrt{\lambda_H \lambda_L} \text{Re}(g_0 \,\tilde{h}_+^{*} ) \nonumber \\
     & + \text{Re}(C_{9-}^{\ell_1 \ell_2}\, C_{T5}^{*\ell_1 \ell_2}) (m_{\ell_1} -m_{\ell_2}) (\mLamB + \mLam) \sqrt{\lambda_H \lambda_L} \big[ \text{Re}(g_0 \tilde{h}_+^{*}) + 2\text{Re}( g_\perp h_\perp^{*}) \big] \nonumber \\
     & + \text{Re}(C_{10+}^{\ell_1 \ell_2}\, C_{T5}^{*\ell_1 \ell_2}) (m_{\ell_1} +m_{\ell_2}) (\mLamB - \mLam) \sqrt{\lambda_H \lambda_L} \big[ \text{Re}(f_0 h_+^{*}) + 2\text{Re}( f_\perp \tilde{h}_\perp^{*}) \big] \nonumber \\ 
     & +q^2 \frac{(\mLamB + \mLam)}{(m_b + m_s)} \text{Re}(C_{S-}^{\ell_1 \ell_2} C_{T5}^{*\ell_1 \ell_2}) \sqrt{\lambda_H \lambda_L} \text{Re}(g_0 \tilde{h}_+^{*}) + q^2 \frac{(\mLamB - \mLam)}{(m_b -m_s)}\text{Re}(C_{P+}^{\ell_1 \ell_2} C_{T5}^{*\ell_1 \ell_2}) \sqrt{\lambda_H \lambda_L} \text{Re}(f_0 h_+^{*}) \bigg\}   \,,\label{eq:b}\\
    c =&+ (|C^{\ell_1\ell_2}_{9+}|^2+|C^{\ell_1\ell_2}_{10+}|^2) \frac{\lambda_L\lambda_H}{q^2 s_+}\bigg[-|f_+|^2 \frac{(\mLamB+\mLam)^2}{q^2}+|f_\perp|^2\bigg] \nonumber\\
    &+ (|C^{\ell_1\ell_2}_{9-}|^2+|C^{\ell_1\ell_2}_{10-}|^2) \frac{\lambda_L\lambda_H}{q^2 s_-}\bigg[-|g_+|^2 \frac{(\mLamB-\mLam)^2}{q^2 }+|g_\perp|^2\bigg]\nonumber \\
    &+(|C^{\ell_1\ell_2}_{T}|^2+|C^{\ell_1\ell_2}_{T5}|^2)\frac{4\lambda_L}{q^2}\bigg[s_- |h_+|^2+s_+|\tilde h_+|^2-\frac{(\mLamB+\mLam)^2s_-}{q^2}|h_\perp|^2-\frac{(\mLamB-\mLam)^2s_+}{q^2}|\tilde h_\perp|^2\bigg]\,,\label{eq:c}
\end{align}
with $C^{\ell_1 \ell_2}_{X \pm} = (C^{\ell_1 \ell_2}_{X} \pm C^{\prime \ell_1 \ell_2}_{X})$, $q_{\pm}= (m_{\ell_{1}}\pm m_{\ell_{2}})^2-q^2$ and $\sigma^{\mu\nu} = i/2 [\gamma^\mu,\gamma^\nu]$. Our results agree with \cite{Boer:2014kda} when setting $m_{\ell_{1}}=m_{\ell_{2}}=0$ and with \cite{Blake:2017une} for $m_{\ell_{1}}=m_{\ell_{2}}$. However, we note that our convention for the helicity angle $\cos \theta_\ell$ has the opposite sign that the one in \cite{Boer:2014kda} . 
We also note that our formulae above disagree with Ref.~\cite{Das:2019omf}, in the specific in terms proportional to $(m_{\ell_1}-m_{\ell_2})^2|C^{\ell_1\ell_2}_{9+}|^2 |f_0|^2$ and $(m_{\ell_1}-m_{\ell_2})^2|C^{\ell_1\ell_2}_{9+}|^2 |g_0|^2$. We ascribe these differences to an incorrect treatment of mass effects in \cite{Das:2019omf}.
Our formulae for the angular coefficients $a$, $b$, $c$ also hold for $\Lambda_b\to\Lambda^*\ell_1^-\ell_2^+$ decays, when setting the additional perpendicular $\Lambda_b\to\Lambda^*$ form factor to zero. This is a reasonable approximation as in the Heavy-Quark-Expansion, this form factor is suppressed by $\Lambda_\mathrm{QCD}/m_b$ \cite{Descotes-Genon:2019dbw,Bordone:2021bop,Das:2020cpv}.\\
In the following, we focus on the branching ratio and forward-backward asymmetry:
\begin{align}
     \frac{d\mathcal{B}^{\ell_1\ell_2}}{d q^2}&= 2\Gamma^{(0)} \tau_{\Lambda
     _b}\left(a  + \frac{c}{3}\right)\,, \label{eq:def_totalrate}\\
     \frac{d A_\mathrm{FB}^{\ell_1\ell_2}}{dq^2} &= \frac{\int_0^1 d\cos\theta\frac{d\Gamma}{d\cos\theta d q^2}-\int_{-1}^0 d\cos\theta\frac{d\Gamma}{d\cos\theta d q^2 }}{\int_0^1 d\cos\theta\frac{d\Gamma}{d\cos\theta d q^2}+\int_{-1}^0 d\cos\theta\frac{d\Gamma}{d\cos\theta d q^2 }} = \frac{b}{2\left(a+\frac{c}{3}\right)}\,,\label{eq:def_AFB}
\end{align}
where the branching ratio $\mathcal{B} = \tau_{\Lambda_b} \Gamma$, where $\tau_{\Lambda_b}$ is the mean life of the $\Lambda_b$ baryon \cite{Amhis:2019ckw} and $\Gamma$ the total width.  
%
To evaluate the size of LFV $\Lambda_b\to\Lambda \ell_1^-\ell_2^+$ decay, we provide the $q^2$-integrated quantities of \eqs{eq:def_totalrate}{eq:def_AFB}. For simplicity, we set $C_i^{\prime\ell_1\ell_2}=0$. We further set $C^{\ell_1\ell_2}_{T(5)}=0$. This choice is discussed at the beginning of \sec{sec:3}. Using the values for the masses from PDG \cite{Zyla:2020zbs}, CKM factors from the UT-fit collaboration \cite{UTFit} and lattice QCD inputs for the form factors\cite{Detmold:2016pkz}, we obtain 
\begin{align}
  10^{8} \cdot \mathcal{B}^{\ell_1\ell_2} & =\, \xi_9^{\ell_1\ell_2} |C_{9}^{\ell_1\ell_2}|^2 +\xi_{10}^{\ell_1\ell_2} |C_{10}^{\ell_1\ell_2}|^2 +\xi_S^{\ell_1\ell_2} |C_{S}^{\ell_1\ell_2}|^2 + \xi_P^{\ell_1\ell_2}|C_{P}^{\ell_1\ell_2}|^2 \nonumber\\ 
   & +\xi_{9S}^{\ell_1\ell_2}\mathrm{Re}( C_{9}^{\ell_1\ell_2} C_{S}^{*\ell_1\ell_2} )  +\xi_{10P}^{\ell_1\ell_2} \, \mathrm{Re}( C_{10}^{\ell_1\ell_2} C_{P}^{*\ell_1\ell_2} )\,,
   \label{eq::Integrated-Br} \\
  A_\mathrm{FB}^{\ell_1\ell_2} &= \bigg[\rho^{\ell_1\ell_2} (|C_{10}^{\ell_1\ell_2}|^2 +|C_{9}^{\ell_1\ell_2}|^2 )+\rho_{910}^{\ell_1\ell_2} \mathrm{Re}( C_{9}^{\ell_1\ell_2} C_{10}^{*\ell_1\ell_2}) \nonumber\\ 
  & +\rho_{9S}^{\ell_1\ell_2}\mathrm{Re}( C_{9}^{\ell_1\ell_2} C_{S}^{*\ell_1\ell_2} ) 
  +\rho_{10P}^{\ell_1\ell_2}\mathrm{Re}( C_{10}^{\ell_1\ell_2} C_{P}^{*\ell_1\ell_2} )\bigg]\bigg/ \Gamma^{\ell_1 \ell_2}\,, 
  \label{eq:results}
\end{align}
with the numerical values for the coefficients $\xi^{\ell_1\ell_2}_i$ and $\rho^{\ell_1\ell_2}_i$ listed in \tables{tab:resultsxi}{tab:resultsrho} and $\Gamma^{\ell_1\ell_2}$ is the integrated width. We present only explicit results for the final states $\tau^{\pm}\mu^{\mp}$ and $\mu^{\pm}e^{\mp}$. The results for $\tau^\pm e^{\mp}$ can easily be obtained from the above results and agree within 1$\sigma$ with those of $\tau^\pm \mu^\mp$. The quoted uncertainties only include those from the form factors, which are the dominant ones. The correlation matrices between the two sets of $\lbrace\xi^{\ell_1\ell_2}_i,\rho^{\ell_1\ell_2}_i\rbrace$ coefficients are given in Appendix~\ref{sec:correlations}. The $\xi_i^{\ell_1\ell_2}$ coefficients in Table~\ref{tab:resultsxi} do not depend on the charges of the final state leptons, except for $ \xi^{\mu\tau}_{9S}$ which depends on $(m_{\ell_{1}}-m_{\ell_{2}})$ and thus switches sign when switching the charges of the final state leptons, i.e. $\xi^{\tau\mu}_{9S} = - \xi^{\mu\tau}_{9S}$. Besides, we note that for $\mu e$ final states, $\xi_9^{\ell_1\ell_2} = \xi^{\ell_1\ell_2}_{10}$ and $\xi_S^{\ell_1\ell_2}=\xi_P^{\ell_1\ell_2}$, such that only the combination $|C_9^{\ell_1\ell_2}|^2 + |C_{10}^{\ell_1\ell_2}|^2$ and $|C_P^{\ell_1\ell_2}|^2 + |C_{S}^{\ell_1\ell_2}|^2$ can be constrained. The coefficients $\rho_i^{\ell_1\ell_2}$ in Table~\ref{tab:resultsrho} are reported in units of $10^{-21}$ GeV$^{-1}$, which is then compensated in $A_\mathrm{FB}^{\ell_1\ell_2}$ by the size of the decay width.

\begin{table}[t]
\begin{center}
\renewcommand{\arraystretch}{1.2} 
\begin{tabular}{c c  c }
\toprule
 & $\ell_1 = \mu,\,\ell_2 = \tau$ & $\ell_1 = \mu,\,\ell_2 = e$  \\
\midrule
$\xi^{\ell_1\ell_2}_{9}$ & $2.15\pm 0.11$  & $3.13\pm0.20$ \\
$\xi^{\ell_1\ell_2}_{10}$  & $2.08\pm0.10$  & $3.13\pm0.20$  \\
$\xi^{\ell_1\ell_2}_{S}$& $0.980\pm0.057$ &  $1.83\pm0.11$  \\
$\xi^{\ell_1\ell_2}_{P}$ & $1.06\pm0.06$ &  $1.83\pm0.11$  \\
$\xi^{\ell_1\ell_2}_{9S}$ & $-0.973\pm0.059$ & $0.142\pm 0.013$ \\
$\xi^{\ell_1\ell_2}_{10P}$ & $1.20\pm0.07$ & $0.144\pm0.013$  \\
\toprule
\end{tabular}
\caption{Numerical values for the parameters of Eq.~\eqref{eq::Integrated-Br}. The coefficients do not depend on the charges of the final state leptons, except for $\xi^{\ell_1\ell_2}_{9S}$ which changes sign when switching the charges of the leptons, i.e. $\xi^{\tau\mu}_{9S} = - \xi^{\mu\tau}_{9S}$ and $\xi^{e \mu}_{9S} = - \xi^{\mu e}_{9S}$. The uncertainties only include those from the form factor which are the dominant ones.}
\label{tab:resultsxi}
\end{center}
\end{table}
\begin{table}[t]
\begin{center}
\renewcommand{\arraystretch}{1.2} 
\begin{tabular}{c c c c c }
\toprule
 & $\ell_1 = \mu,\,\ell_2 = \tau$ & $\ell_1 = \tau,\,\ell_2 = \mu$ & $\ell_1 = \mu,\,\ell_2 = e$ & $\ell_1 = e,\,\ell_2 = \mu$ \\
\midrule
$\rho^{\ell_1\ell_2}$ & $\phantom{-}1.26\pm0.08$ & $-1.26\pm0.08$  &  $-0.025\pm0.005$ & $\phantom{-}0.025\pm0.005$ \\
$\rho^{\ell_1\ell_2}_{910}$ & $-5.09\pm0.24$  & $-5.09\pm0.24$ & $-9.16\pm0.55$ & $-9.16\pm0.55$  \\
$\rho^{\ell_1\ell_2}_{9S}$ & $-2.23\pm0.12$& $-2.23\pm0.12$ & $-0.283\pm0.023$ & $-0.283\pm0.023$  \\
$\rho^{\ell_1\ell_2}_{10P}$ & $\phantom{-}1.99\pm0.11$ & $-1.96\pm0.11$ & $-0.280\pm0.023$ & $\phantom{-}0.280\pm0.023$ \\ 
\toprule
\end{tabular}
\caption{Coefficients for the numerator of $A_\mathrm{FB}^{\ell_1
\ell_2}$. We give the values in units of $10^{-21} \,\, \rm{GeV}^{-1}$. This factor is compensated by the size of the decay width in \Eq{eq:results}. 
}
\label{tab:resultsrho}
\end{center}
\end{table}

\section{Phenomenological implications}
\label{sec:3}
In the following, we discuss the implications of the available constraints on LFV $B$-meson decays and which bounds they imply on the observables in the baryonic modes. In order to do so, we need to choose which NP operators are present. Since no NP particles have been observed so far above the electroweak scale, we choose to work with the SMEFT:
\begin{equation}
\begin{aligned}
\mathcal{L}_\text{eff}=\mathcal{L}_\text{SM} - \frac{1}{M^2}\bigg\{[&\mathcal{C}_{lq}^{(3)}]^{ij\alpha\beta}(\bar{Q}^i \gamma^\mu \sigma^a Q^j)(\bar{L}^\alpha \gamma_\mu \sigma^a L^\beta)+[\mathcal{C}_{lq}^{(1)}]^{ij\alpha\beta}(\bar{Q}^i \gamma^\mu Q^j)(\bar{L}^\alpha \gamma_\mu  L^\beta) \\
+&[\mathcal{C}_{ledq}]^{ij\alpha\beta}(\bar{Q}^i d_R^j)(\bar{e}_R^\alpha L^\beta) \bigg\} \,,
\end{aligned}
\label{eq:lagrangian_SMEFT}
\end{equation}
where we adopt the so-called Warsaw basis \cite{Grzadkowski:2010es}. Here we denote with $Q$ and $L$ the left-handed quark and lepton doublets, respectively, and with $e_R$ and $d_R$ the right-handed charged leptons and down-type quarks, respectively. We further denote $\epsilon = i\sigma_2$ and $M$ is the effective scale which can be associated with the mass of the heavy NP degrees of freedom.\\
The operators in \Eq{eq:lagrangian_SMEFT} are the complete set of dimension-6 semileptonic operators that can contribute to $b\to s\ell_1\ell_2$ transitions. We note that none of these operators contain a tensor current; nonetheless, at low energy, the operator $\mathcal{O}_{T(5)}^{\ell_1\ell_2}$ defined in \Eq{eq:ops} could be generated through effective operators containing a covariant derivative \cite{Bobeth:2007dw}. However, as tensor operators provide a poor explanation for $B$ anomalies (see e.g. \cite{Hiller:2014yaa}), we do not consider them in our analysis. 

The Wilson coefficients of the operators in \Eq{eq:lagrangian_SMEFT} can be constrained from low-energy processes as well as high-$p_T$ data, and in general a flavour structure has to be assumed to reduce the number of independent NP parameters. In the following, we choose to consider only constraints from low-energy data and first do not to assume any hierarchy for the NP couplings. In \sec{sec:3.2} we then study particular scenarios, where a more complex structure for NP couplings in flavour space is assumed. For the $b\to s \ell_1 \ell_2$ transition we are interested in, we set $i=2$ and $j=3$ and generic $\alpha=\ell_1$ and $\beta=\ell_2$ in \Eq{eq:lagrangian_SMEFT}. Performing now the tree-level matching onto \Eq{eq:heff}, we have
\begin{equation}
\begin{aligned}
C_{9}^{\ell_1 \ell_2} = - \,{C}_{10}^{\ell_1 \ell_2} =& +\frac{v^2}{\Lambda^2}\frac{\pi}{\alpha_\text{em} |V_{tb}V^*_{ts}|} \left([\mathcal{C}_{lq}^{(3)}]^{23\ell_1 \ell_2}+[\mathcal{C}_{lq}^{(1)}]^{23\ell_1 \ell_2}\right)\,, \\
C_{9}^{\prime\,\ell_1 \ell_2} = +\, {C}_{10}^{\prime\, \ell_1\ell_2} = &+ \frac{v^2}{\Lambda^2}\frac{\pi}{\alpha_\text{em} |V_{tb}V^*_{ts}|}  [\mathcal{C}_{l d}]^{23\ell_1 \ell_2}\,, \\
{C}_S^{\ell_1 \ell_2} = -\,{C}_{P}^{\ell_1 \ell_2} = & +\frac{v^2}{\Lambda^2}\frac{\pi}{\alpha_\text{em} |V_{tb}V^*_{ts}|} \,[\mathcal{C}_{leqd}]^{23\ell_1 \ell_2}\,, \\
{C}_S^{\prime\,\ell_1 \ell_2} = +\,{C}_{P}^{\prime\, \ell_1\ell_2} = & + \frac{v^2}{\Lambda^2}\frac{\pi}{\alpha_\text{em} |V_{tb}V^*_{ts}|} \,[\mathcal{C}_{leqd}^*]^{32\ell_1 \ell_2}\,.
\end{aligned}
\label{eq:observables_bsll}
\end{equation}

\subsection{Model-independent approach}
\label{sec:3.1}
\begin{table}[t]
\begin{center}
\renewcommand{\arraystretch}{1.3} 
\begin{tabular}{c c}
\toprule
Observable & Upper Bound \\
\midrule
$\mathcal{B}(\bar{B}_s \to \mu^{\pm} \tau^{\mp})$ & $ 3.5
\cdot 10^{-5}$\, \cite{Aaij:2019okb} \\
$\mathcal{B}(\bar B_s\to\mu^\pm e^\mp)$ & $5.4\cdot 10^{-9}$ \cite{Aaij:2017cza} \\ 
$\mathcal{B}(B^+ \to K^+  \tau^{-}\mu^{+})$ &  $4.5 \cdot 10^{-5}$\, \cite{Lees:2012zz} \\
$\mathcal{B}(B^+\to K^+  \mu^{-}\tau^{+})$ &  $3.9 \cdot 10^{-5}$\, \cite{Aaij:2020mqb} \\ 
$\mathcal{B}(B^+ \to K^+ \mu^{-} e^{+} )$ &  $7.0 \cdot 10^{-9}$\, \cite{Aaij:2019nmj}  \\
$\mathcal{B}(B^+ \to K^+ e^{-} \mu^{+})$ &  $6.4 \cdot 10^{-9}$\, \cite{Aaij:2019nmj}  \\
\toprule
\end{tabular}
\caption{Experimental upper limits for LFV $B$ decays at 90$\%$ C.L..} 
\label{tab:UpperBounds-inputs}
\end{center}
\end{table}
\begin{table}[t]
\begin{center}
\renewcommand{\arraystretch}{1.2} 
\begin{tabular}{c  c c }
\toprule
 & $\ell_1^- = \mu^-,\,\ell_2^+ = \tau^+$  & $\ell_1^- = \mu^-,\,\ell_2^+ = e^+$  \\
\midrule
$c_{\ell_1\ell_2}^{9+}$ & $\phantom{-}1.09$
& $1.75$ \\ 
$c_{\ell_1\ell_2}^{10+}$ & $\phantom{-}1.14$
& $1.75$ \\ 
$c_{\ell_1\ell_2}^{S}$ & $\phantom{-}1.47$
& $2.68$\\ 
$c_{\ell_1\ell_2}^{P}$ & $\phantom{-}1.58$
& $2.68$  \\  
$c_{\ell_1\ell_2}^{S9}$ & $-1.35$
& $0.21$ \\ 
$c_{\ell_1\ell_2}^{P10}$ & $\phantom{-}1.66$
   &$0.21$  \\ 
\toprule
\end{tabular}
\caption{Predictions for the coefficients describing $B^+\to K^+ \ell_1^-\ell_2^+$ decays using the hardonic form factors from Ref.~\cite{Bouchard:2013eph,Gubernari:2018wyi}. We note that these coefficients are independent of the charges of the leptons, except for  $c_{\ell_1\ell_2}^{S9}$ which changes sign depending on the charge of the heavier lepton. }
\label{tab:BtoKtaumu}
\end{center}
\end{table}
First, we consider the constraints on several combinations of Wilson coefficients from measurements of mesonic LFV decays. We consider the branching ratios of the decay modes $\bar{B}_s \to\ell_1^-\ell_2^+$ and $B\to K \ell_1^-\ell_2^+$, for which the experimental upper limits at $90\%$ C.L. are reported in \Table{tab:BtoKtaumu}. Using \Eq{eq:heff}, we have: 
\begin{equation}
\begin{aligned}
\mathcal{B}(\bar{B}_s\to\ell^-_1 \ell^+_2)=&\,\frac{\tau_{B_s}}{64\pi^3}\frac{\alpha_\text{em}^2G_F^2 |V_{tb}V_{ts}^*|^2}{m_{B_s}^3} f_{B_s}^2 \, \lambda^{1/2}(m_{B_s}^2,m_{\ell_1}^2,m_{\ell_2}^2) \times \\
\times&\left\{[m_{B_s}^2-(m_{\ell_1}-m_{\ell_2})^2]\left|(m_{\ell_1}+m_{\ell_2}){C}^{\ell_1\ell_2}_{10-}+\frac{m_{B_s}^2}{m_b+m_s}{C}^{\ell_1\ell_2}_{P-}\right|^2\right. \\
&+\left.[m_{B_s}^2-(m_{\ell_1}+m_{\ell_2})^2]\left|(m_{\ell_1}-m_{\ell_2})({C}^{\ell_1\ell_2}_{9-})+\frac{m_{B_s}^2}{m_b+m_s}({C}^{\ell_1\ell_2}_{S-})\right|^2\right\} \,,
\label{eq::Br-Bs2ll}
\end{aligned}
\end{equation}
and
\begin{equation}
\begin{aligned}
\mathcal{B}(B^+\to K^+\ell_1^-\ell_2^+) &= 10^{-8} \bigg\{c^{9+}_{\ell_1\ell_2}\left|C_{9+}^{\ell_1\ell_2} \right|^2 + c^{10+}_{\ell_1\ell_2}\left|C_{10+}^{\ell_1\ell_2} \right|^2  +  c^S_{\ell_1\ell_2}\left|C_{S+}^{\ell_1\ell_2} \right|^2 \\[2pt]
 &+c^P_{\ell_1\ell_2}\left|C_{P+}^{\ell_1\ell_2}\right|^2 +c^{S9}_{\ell_1\ell_2}\,\mathrm{Re}[C_{S+}^{*\ell_1\ell_2}  C_{9+}^{\ell_1\ell_2}] + c^{P10}_{\ell_1\ell_2}\,\mathrm{Re}[C_{P+}^{*\ell_1\ell_2} C_{10+}^{\ell_1\ell_2}]
 \bigg\}\,, 
 \label{eq::Br-B2Kll}
\end{aligned}
\end{equation}
Both \eqs{eq::Br-Bs2ll}{eq::Br-B2Kll} agree with previous results in the literature \cite{Becirevic:2016oho,Gratrex:2015hna}.
Using again the values for the masses from PDG \cite{Zyla:2020zbs}, CKM factors from the UT-fit collaboration \cite{UTFit}, $f_{B_s}=215\,\mathrm{MeV}$ \cite{Balasubramanian:2019net} and Lattice QCD/Light Cone Sum Rule results in Refs.~\cite{Bouchard:2013eph,Gubernari:2018wyi}, we find the coefficients $c_{\ell_1\ell_2}^i$ in \Eq{eq::Br-B2Kll} as listed in \Table{tab:BtoKtaumu}. Similar as for $\xi_{9S}$, the coefficient $c_{\ell_1\ell_2}^{S9}$ is proportional to $m_{\ell_1}-m_{\ell_2}$ and thus changes sign depending on charge of the heavier lepton. We stress that the numbers in \Table{tab:BtoKtaumu} are strongly dependent on the choice for $\alpha_\mathrm{em}$. Here we take $\alpha_\mathrm{em}=1/133$. A different choice can be implemented by rescaling the $c_{\ell_1\ell_2}^i$ coefficients.

Finally, we use the experimental upper bounds listed in \Table{tab:UpperBounds-inputs} and \eqs{eq::Br-Bs2ll}{eq::Br-B2Kll} to constrain different combinations of couplings $C_{i}^{\ell_1, \ell_2}$. As stated before, we do not consider $\tau e$ decays as the constraints coming from these decays are similar to those from the $\tau \mu$ channel. Furthermore, for simplicity we also do not consider the $\mathcal{O}^{\prime\ell_1\ell_2}_i$ operators. This choice is motivated by the fact that these operators are unappealing when trying to fit $b\to s \ell \ell$ data \cite{Alguero:2021anc,Altmannshofer:2021qrr,Hurth:2021nsi,Ciuchini:2019usw,Lancierini:2021sdf}. Nevertheless, we stress that the baryonic channels have a different dependence on the primed operators  with respect to the mesonic ones, which may be interesting to consider once scenarios involving these operators become more interesting to explain the $B$ anomalies. 

\begin{figure}[t]
	\centering
  	\subfloat{\includegraphics[width=0.5\textwidth]{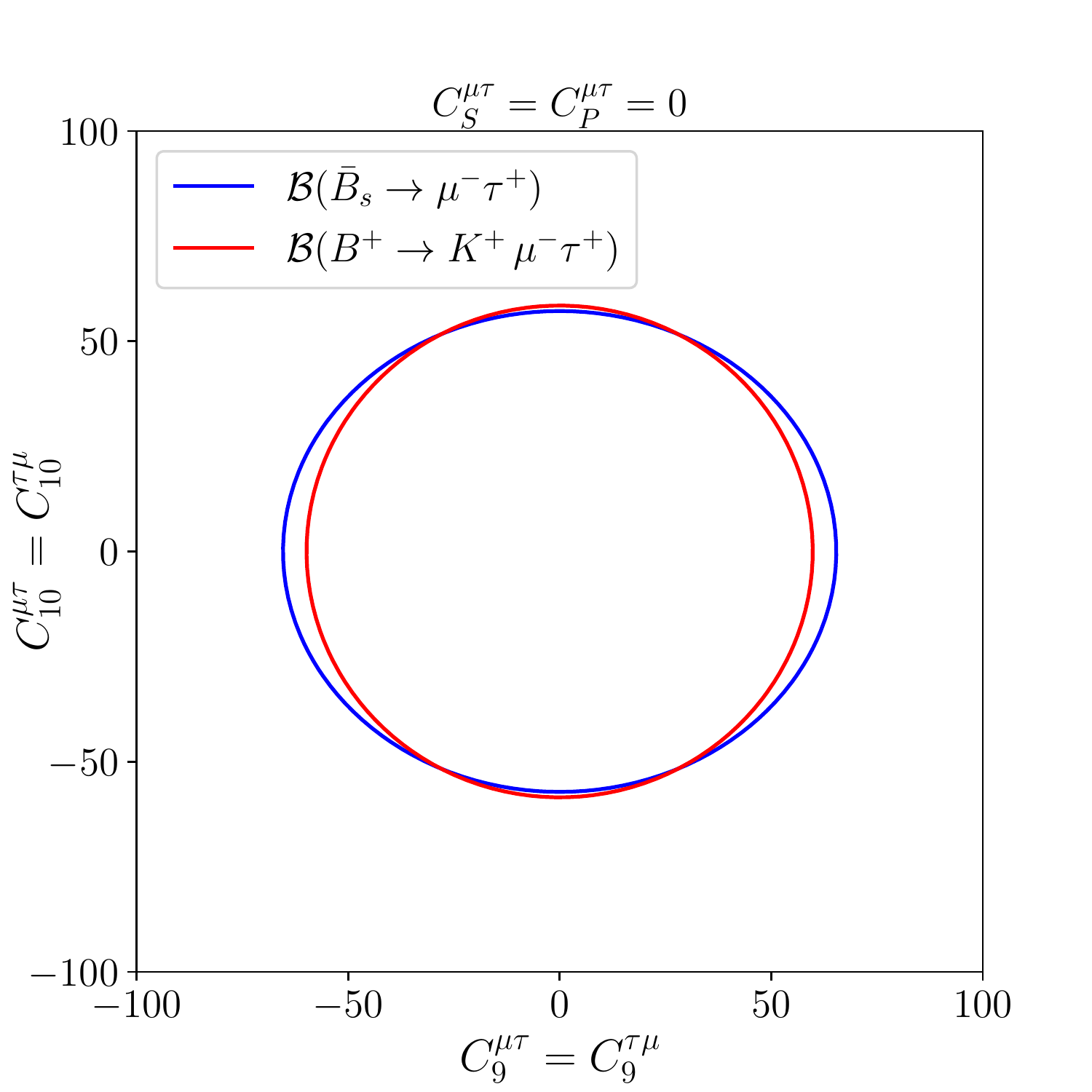}} 
  	\subfloat{\includegraphics[width=0.5\textwidth]{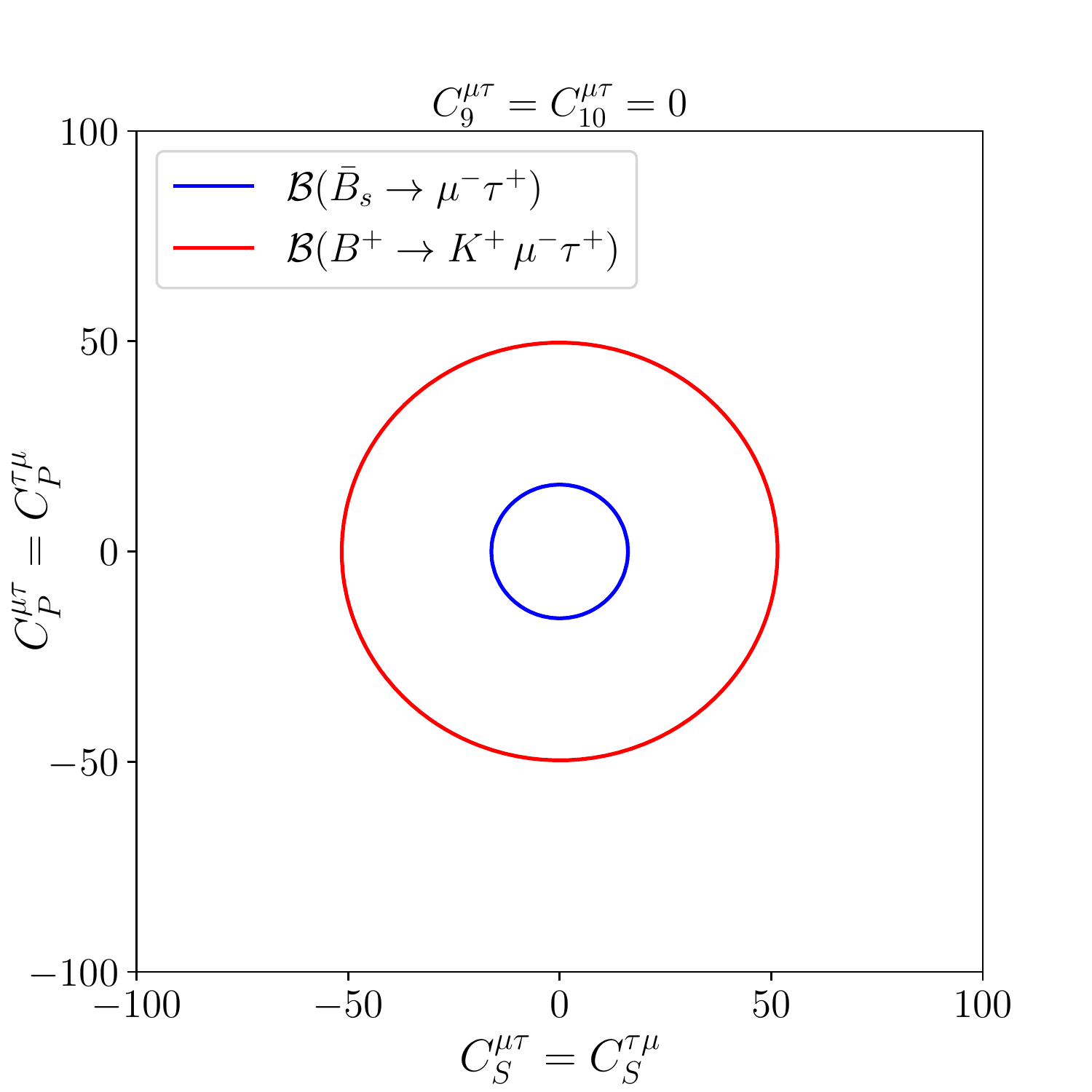}} \\
  	\subfloat{\includegraphics[width=0.5\textwidth]{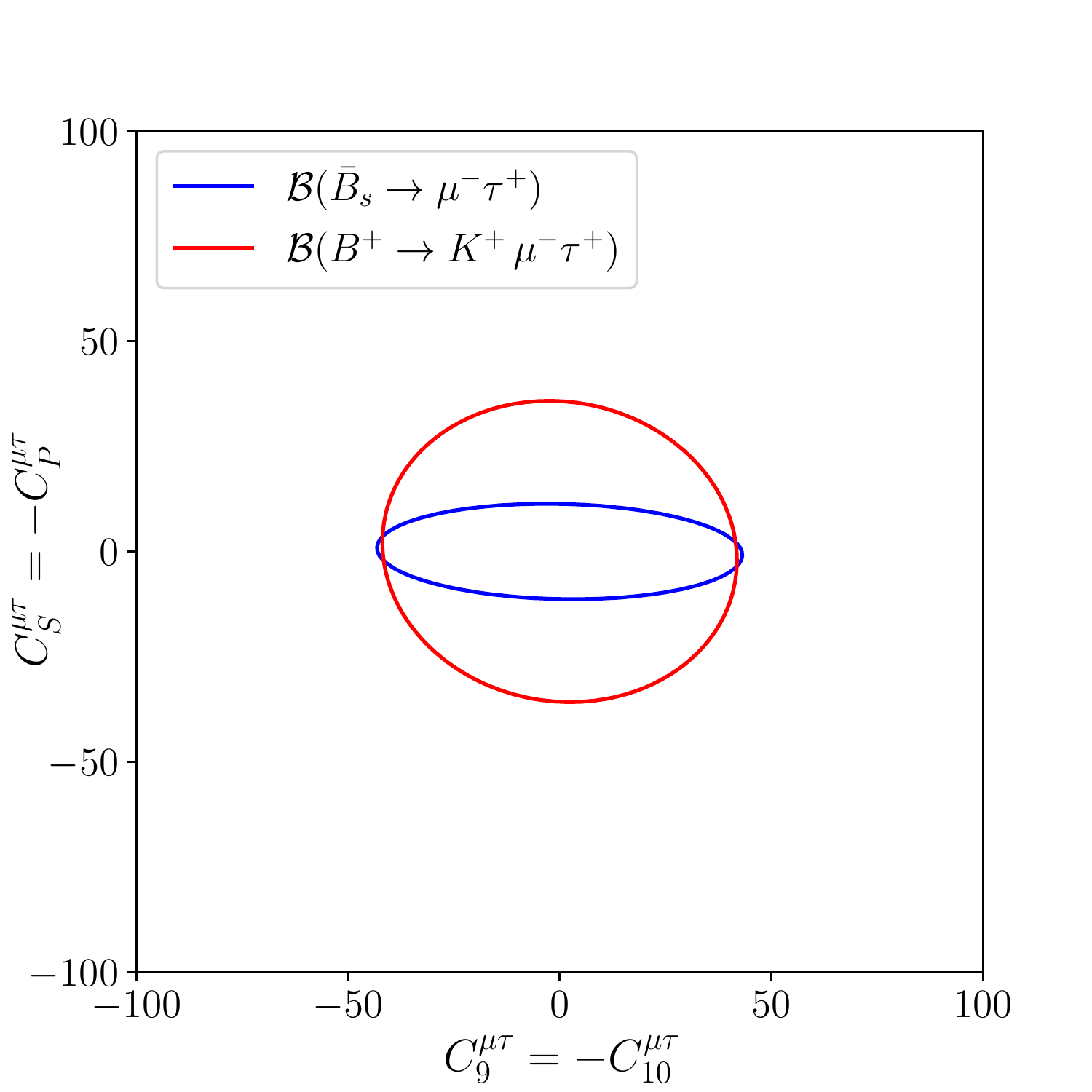}} 
  	  	\subfloat{\includegraphics[width=0.5\textwidth]{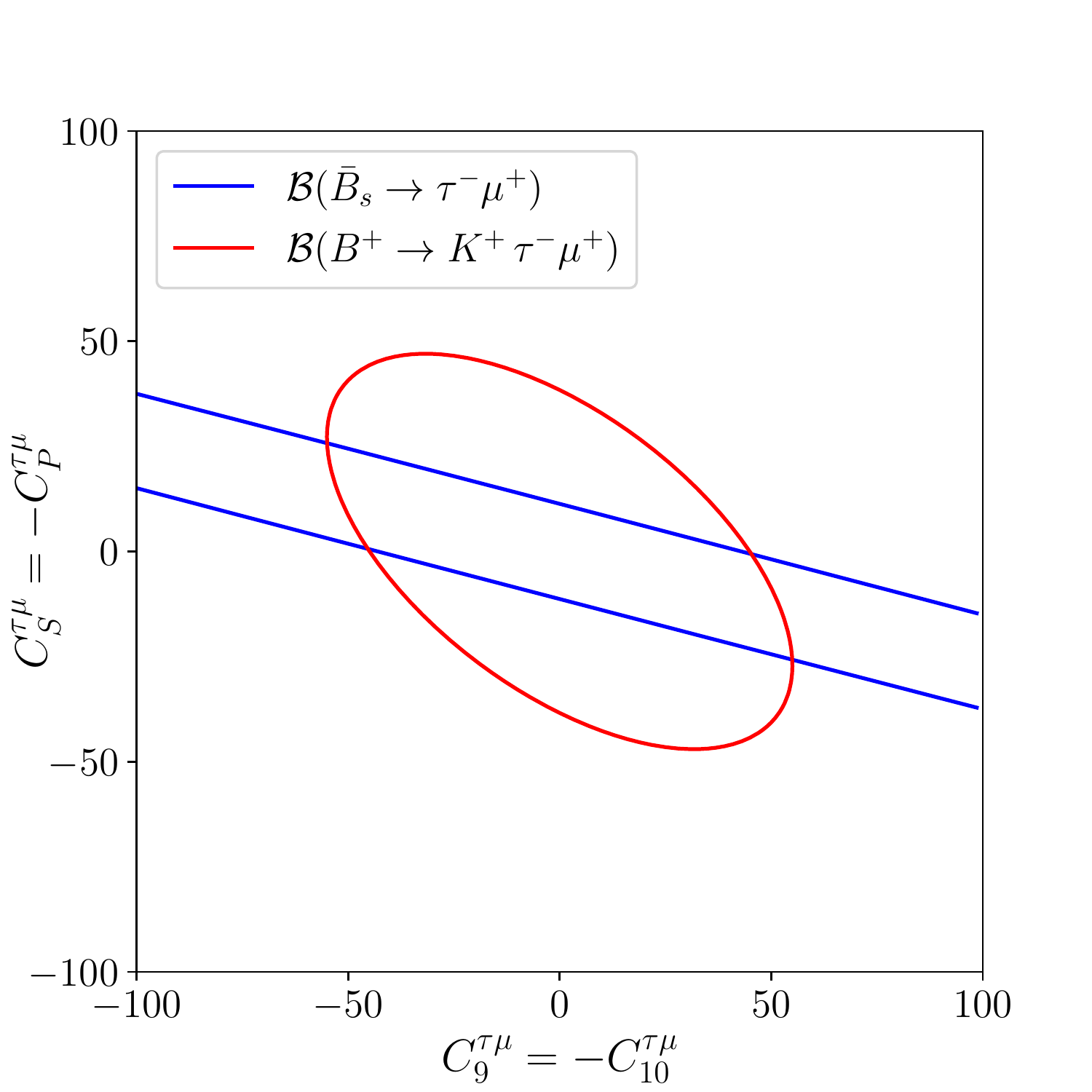}} 
    \caption{Model independent constrains on different combinations of Wilson coefficients obtained from the $90\%$ C.L. upper limits on meson $b\to s \mu^\pm \tau^\mp$ transitions.}
	\label{fig::ModelIndep-constrainstau}
\end{figure}

\begin{figure}[t]
	\centering
	  	\subfloat{\includegraphics[width=0.5\textwidth]{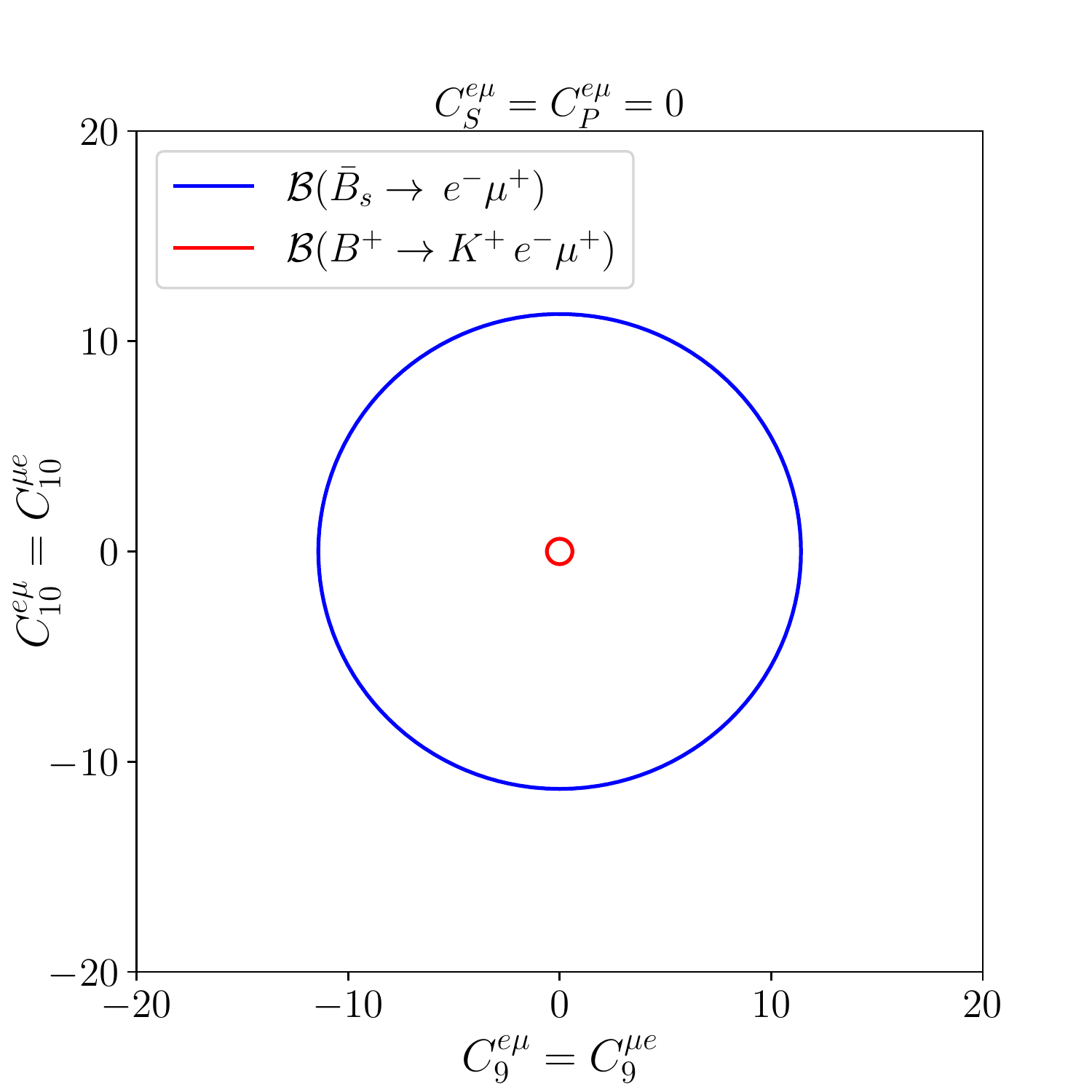}} 
	\subfloat{\includegraphics[width=0.5\textwidth]{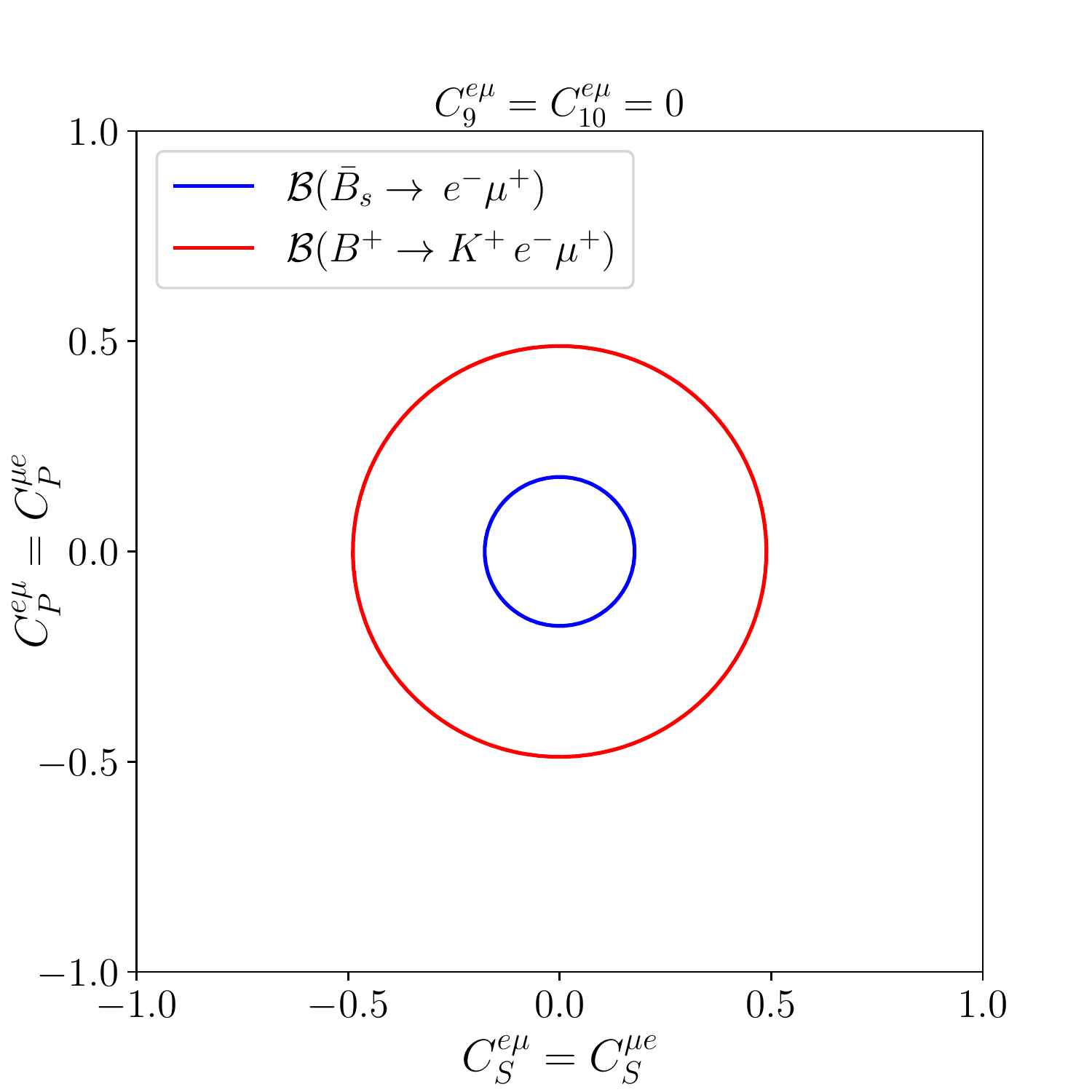}} \\
  	\subfloat{\includegraphics[width=0.5\textwidth]{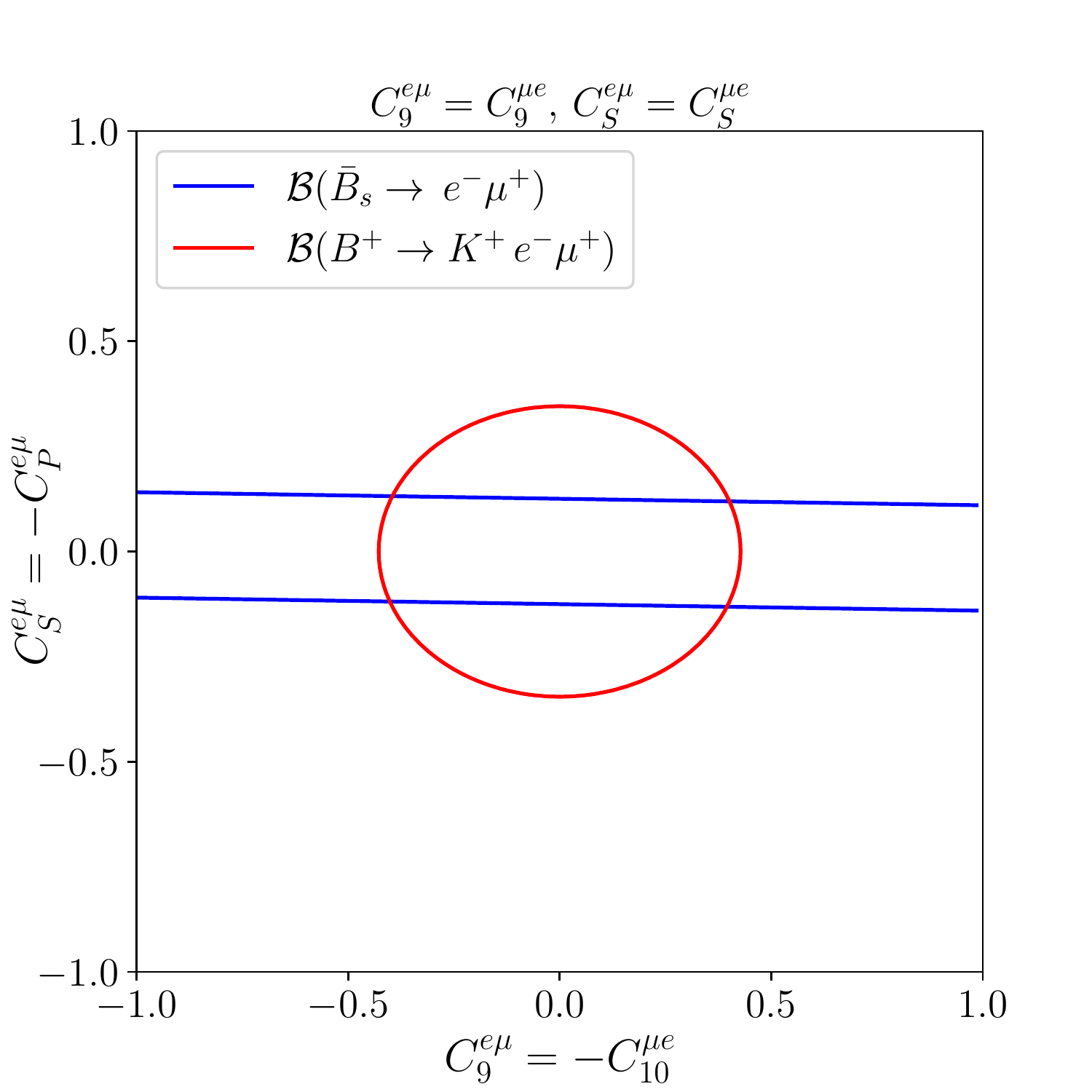}} 
    \caption{Model independent constrains on different combinations of Wilson coefficients obtained from the $90\%$ C.L. upper limits on meson $b\to s \mu^\pm e^\mp$ transitions.}
	\label{fig::ModelIndep-constrainsel}
\end{figure}
The obtained bounds for $\tau\mu$ and $\mu e$ finals states are given in Fig.~\ref{fig::ModelIndep-constrainstau} and in Fig.~\ref{fig::ModelIndep-constrainsel}, respectively.
We consider three 2-dimensional scenarios, in which we allow only some combinations of NP Wilson coefficients to be non-zero: $C_9^{\ell_1\ell_2}$ and $C_{10}^{\ell_1\ell_2}$, $C_S^{\ell_1\ell_2}$ and $C_P^{\ell_1\ell_2}$ and the SMEFT inspired one, where $C_9^{\ell_1\ell_2}=-C_{10}^{\ell_1\ell_2}$ and $C_S^{\ell_1\ell_2}=-C_P^{\ell_1\ell_2}$.
For the $C_9^{\ell_1\ell_2}-C_{10}^{\ell_1\ell_2}$ and $C_S^{\ell_1\ell_2}-C_P^{\ell_1\ell_2}$ scenarios, which are independent of the charge configuration in the final state, we only consider the strongest bound in Table~\ref{tab:UpperBounds-inputs}. As the interference between $C_9^{\ell_1\ell_2}$ and  $C_S^{\ell_1\ell_2}$ depends on the charge configuration of the leptons in the final state, we present plots for both the $\tau^+\mu^-$ and $\tau^-\mu^+$ final states. We note that the $\bar{B}_s\to \tau^- \tau^+$ decay only gives a very weak constraint in the $C_9^{\ell_1\ell_2}=-C_{10}^{\ell_1\ell_2}$ plane ranging from $-200$ to $200$. From comparison of the plots in \fig{fig::ModelIndep-constrainstau}, we find large differences between the $\tau^+\mu^-$ and the $\tau^-\mu^+$. Hence, we stress that it is important to analyse these final states separately. For the electron, the differences between $\mu^-e^+$ and $\mu^+e^-$ are negligible and we only present one figure.   

As the mesonic $B^+\to K^+\ell_1\ell_2$ and $\bar B_s\to \ell_1\ell_2$ are mediated by the same quark level transition, we can use the obtained upper limits on combinations of Wilson coefficients and convert those into upper limit on the branching ratio and forward-backward asymmetry for $\Lambda_b\to \Lambda\ell_1\ell_2$ decays using \Eq{eq:results}. When allowing for only one NP Wilson coefficient to be nonzero at a time, for example allowing only $C_9^{\ell_1\ell_2}\neq 0$, the corresponding bounds can be easily obtained by calculating the scale factor between $c^i_{\ell_1\ell_2}$ of the meson $B\to K$ LFV decay and $\xi_i^{\ell_1\ell_2}$ of the baryon $\Lambda_b\to \Lambda$ decay using \Table{tab:resultsxi} and \Table{tab:BtoKtaumu} and re-scaling the upper limit of the mesonic decay accordingly. In addition, comparing the coefficients in these Tables, we observe that the ratios $c^{i}_{\ell_1\ell_2}/c^{j}_{\ell_1\ell_2}$ and $\xi_{i}^{\ell_1\ell_2}/\xi_{j}^{\ell_1\ell_2}$ are very similar for $i,j=9,10$ and $i,j=S,P$.
Therefore, the sensitivities for LFV $B\to K$ and $\Lambda_b\to\Lambda$ decays are rather similar when considering the $C_9^{\ell_1\ell_2}-C_{10}^{\ell_1\ell_2}$ only and $C_S^{\ell_1\ell_2}-C_P^{\ell_1\ell_2}$ only scenarios. 
Upper limits (at $90\%$ C.L.) for the branching ratio of $\Lambda_b\to \Lambda \ell_1\ell_2$ derived from their mesonic counter parts for the three scenarios are presented in Table~\ref{tab:resultsmodelindep}. These values should be interpreted as follows: any future experimental upper limit on the baryonic mode below the quoted value gives stronger constraints on the Wilson coefficients than those obtained from the current mesonic upper limits.  

\begin{table}[t]
\begin{center}
\renewcommand{\arraystretch}{1.2} 
\begin{tabular}{c c c }
\toprule
 & $ \mathcal{B}^{\mu\tau} \; (\mathcal{B}^{\tau \mu}) \times 10^{-5}$  & $\mathcal{B}^{e \mu} = \mathcal{B}^{\mu e} \times  10^{-8}$\\
\midrule 
$C_9^{\ell_1 \ell_2}\neq 0, C_{10}^{\ell_1 \ell_2} \neq 0, C_S^{\ell_1 \ell_2} =C_P^{\ell_1 \ell_2} =0$ & $ <7.7\; (7.7) $  & $<1.1$ \\
$C_S^{\ell_1 \ell_2} \neq 0, C_{P}^{\ell_1 \ell_2}\neq 0, C_9^{\ell_1 \ell_2}=C_{10}^{\ell_1 \ell_2}=0$ & $<2.7\; (2.7)$  & $<0.06  $  \\
$C_9^{\ell_1 \ell_2}=-C_{10}^{\ell_1 \ell_2}, C_{S}^{\ell_1 \ell_2}=-C_P^{\ell_1 \ell_2}$ & $<7.4 \; (11) $ & $<1.1$  \\
\toprule
\end{tabular}
\caption{Upper limits for the branching ratio of $\Lambda_b \to \Lambda$ LFV decays obtained in a model independent way by considering their mesonic counter parts. Bounds are at $90\%$ C.L.. For the first two scenarios, the branching ratios are independent of the charge configuration. However for the SMEFT scenario this is not the case anymore, hence we present both branching ratios for $\mu^-\tau^+$ and in brackets $\tau^-\mu^+$.}
\label{tab:resultsmodelindep}
\end{center}
\end{table}

\begin{figure}
\centering
\subfloat{\includegraphics[width=0.5\textwidth]{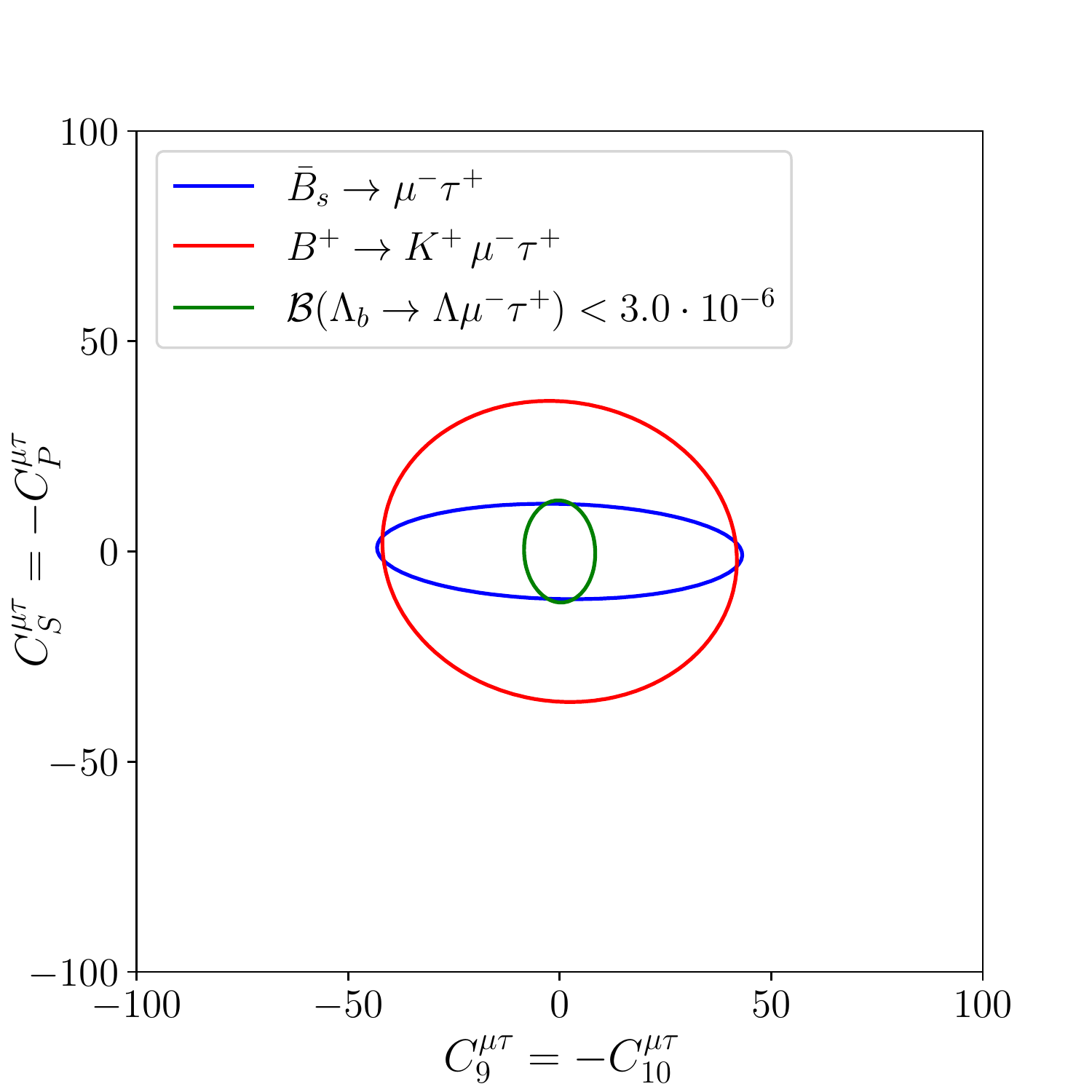}} 
\subfloat{\includegraphics[width=0.5\textwidth]{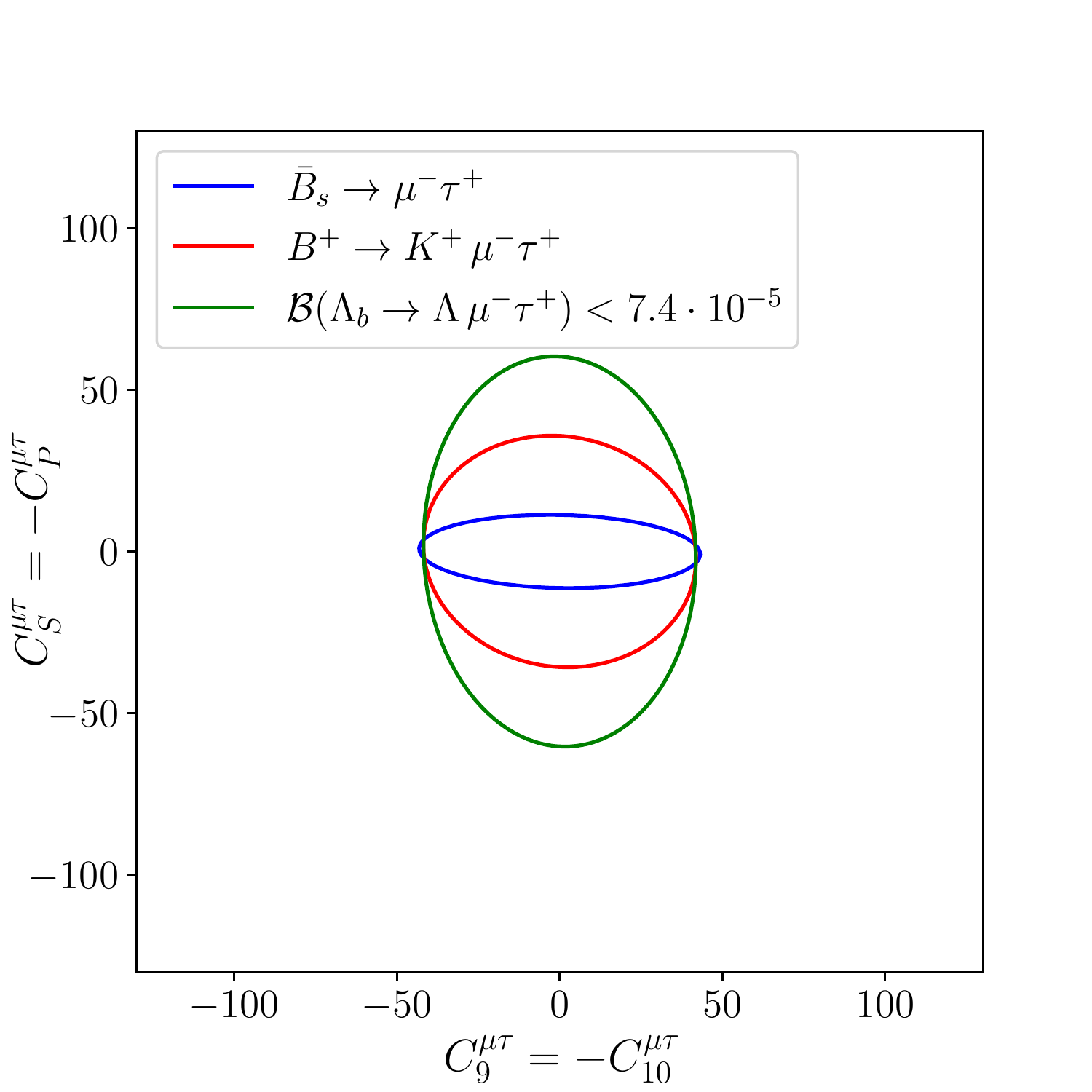}} 
\caption{Illustration of the orthogonality between current mesonic and possible future baryonic constraints. }
\label{fig:ortho}
\end{figure}
The complementarity of the mesonic and the baryonic LFV channels specifically arises when considering both (axial)vector and (pseudo)scalar operators. This complementarity is caused the difference between the ratios  $c^{i}_{\ell_1\ell_2}/c^{j}_{\ell_1\ell_2}$ and $\xi_{i}^{\ell_1\ell_2}/\xi_{j}^{\ell_1\ell_2}$ for $i=S9, P10$ and $j=S,P,9,10$. We expect a similar complementarity also when both tensor operators and (axial)vector operators are present. We illustrate quantitatively this in \fig{fig:ortho} for the SMEFT scenario where $C_9^{\ell_1\ell_2}=-C_{10}^{\ell_1\ell_2}, C_S^{\ell_1\ell_2}=-C_P^{\ell_1\ell_2}$. We present the current meson constraints combined with two possible constraints on the $\Lambda_b \to \Lambda  \mu^-\tau^+$ branching ratio and observe that the mesonic modes place strong constraints on scalar/pseudoscalar interactions while the baryonic channel is more sensitive to $C_9^{\mu\tau}$ and $C_{10}^{\mu\tau}$. 

\begin{figure}
\begin{center}
\centering
\subfloat{\includegraphics[width=0.5\textwidth]{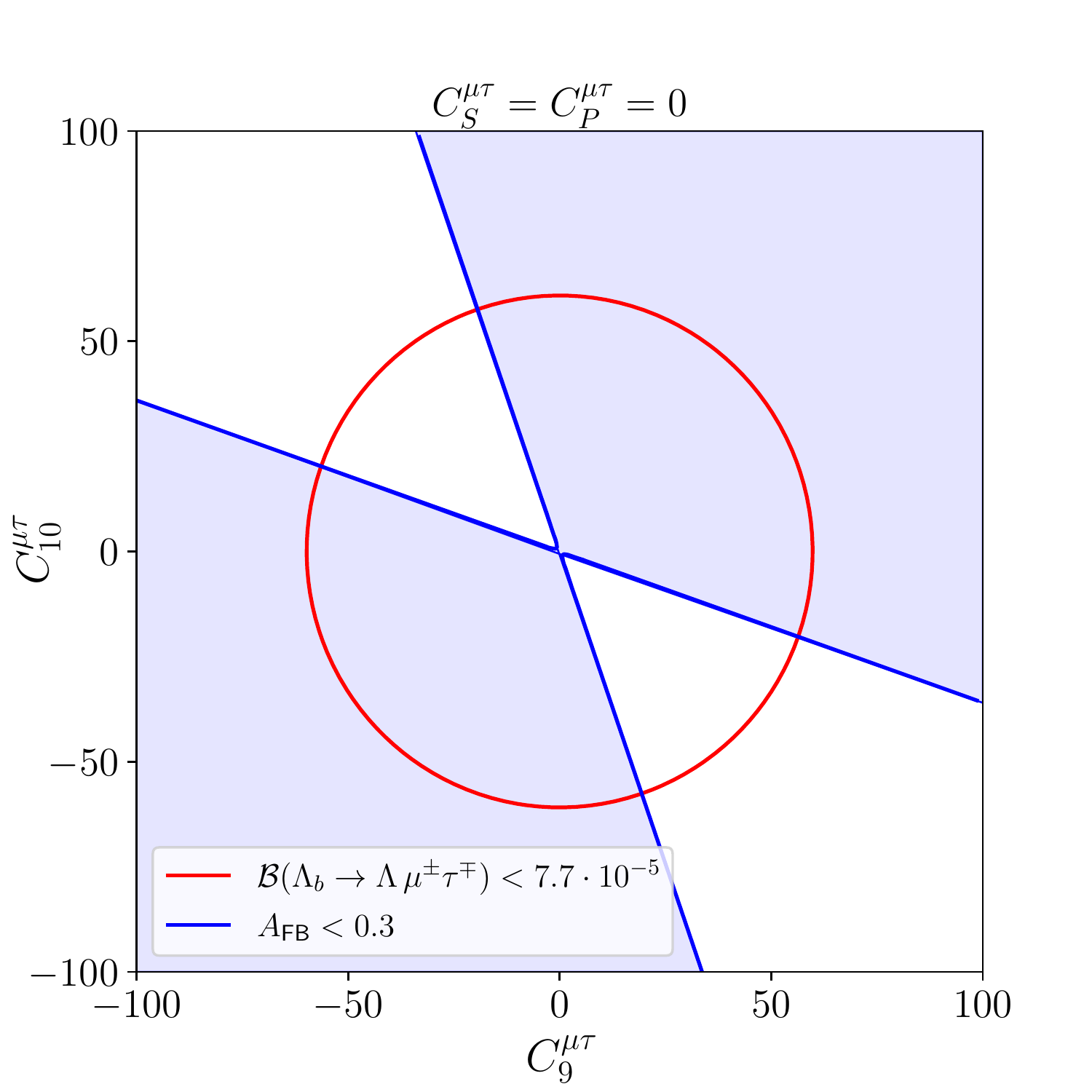}} 
\end{center}
\caption{Illustration of how the forward-backward asymmetry provides orthogonal constraints to the branching ratio of $\Lambda_b\to \Lambda \mu^- \tau^+$. The shaded area present the allowed region for an upper limit $A_{\rm FB}^{\mu\tau}<0.3$, compared to an upper limit for the branching ratio of $7.7\cdot 10^{-5}$.} 
\label{fig:AFB}
\end{figure}

Finally, we consider the integrated forward-backward asymmetry $ A_\mathrm{FB}^{\ell_1\ell_2}$ which provides orthogonal information compared to the branching ratio. From \Eq{eq:results} we note the following properties: $A_\mathrm{FB}^{\ell_1\ell_2}$ is identically zero if $C_9^{\ell_1\ell_2} = C_{10}^{\ell_1\ell_2}=0$, and in the case in which only $C_9^{\ell_1\ell_2} = -C_{10}^{\ell_1\ell_2}\neq0$  $A_\mathrm{FB}^{\ell_1\ell_2}$ is independent on the values of $C_9^{\ell_1\ell_2}$ and $C_{10}^{\ell_1\ell_2}$. In the latter scenario, we find for $ C_9^{\ell_1\ell_2}= - C_{10}^{\ell_1\ell_2}$ and $ C_S^{\ell_1\ell_2} = C_P^{\ell_1\ell_2} =0$
\begin{equation}
\label{eq:afbfix}
\begin{aligned}
     A_\mathrm{FB}^{\tau\mu} = 0.14\pm0.01\ , \quad \quad A_\mathrm{FB}^{\mu\tau} = 0.40\pm0.03\ , \quad \quad  A_\mathrm{FB}^{e\mu} =  A_\mathrm{FB}^{\mu e} = 0.33\pm0.04\ .
\end{aligned}
\end{equation}
A measurement or an upper limit different from these values provides interesting complementary information. This is illustrated in \fig{fig:AFB}, where we consider for the $\mu^- \tau^+$ final state a future scenario in which an upper limit of $A_{\rm FB}^{\mu\tau}<0.3$ and $\mathcal{B}^{\mu \tau} < 7.7\cdot 10^{-5}$ are considered. As we can see from \fig{fig:AFB}, the information on $A_\mathrm{FB}^{\tau \mu}$ helps to rule out a large part of the allowed space in the $C_9^{\ell_1\ell_2}-C_{10}^{\ell_1\ell_2}$ plane.

\subsection{Explicit models}
\label{sec:3.2}
As mentioned, in many models that explain LFU violation, also LFV naturally occurs. Since our aim is not to perform a detailed analysis of all the observables in low-energy phenomenology, we choose to focus here on two specific models that explain the $B$ anomalies. We choose two interesting solutions, which are the most favourite in the literature: the combination of the scalar leptoquarks $S_1$ and $S_3$ and the vector leptoquark $U_1$. For these models we provide predictions for observables in $\Lambda_b\to \Lambda\ell_1^-\ell_2^+$ decays.\\

\newpage\noindent {\bf The \boldmath$S_1+S_3$ scalar leptoquarks scenario {\cite{Bordone:2020lnb}}}\\
Here we focus on the $S_1+S_3$ scenario, following the analysis in Ref.~\cite{Bordone:2020lnb}\footnote{The analysis Ref.~\cite{Bordone:2020lnb} provides qualitatively the same results as Ref.~\cite{Gherardi:2020qhc}.}.
The main idea there is to apply the Froggatt-Nielsen mechanism \cite{Froggatt:1978nt}, that explains the hierarchies of quark masses, as a power counting for NP operators, and thereby providing simultaneously an explanation for the $B$-anomalies and the flavour puzzle. Converting the formalism of Ref.~\cite{Bordone:2020lnb} to the Wilson coefficients defined in \Eq{eq:heff}, we find: 
\begin{equation}
    C_9^{\ell_1\ell_2} = - C_{10}^{\ell_1\ell_2}=\frac{v^2}{M^2} \frac{\pi}{\alpha_\mathrm{em}|V_{tb}V_{ts}^*|} |g_3|^2 \tilde S_{QL}^{3\ell_2}\tilde S_{QL}^{*2\ell_1}    
\end{equation}
where $M$ is the mass of the heavy scalar leptoquarks, $\tilde S_{QL}^{i\ell_i}$ is the spurion associated with the $S_3$ scalar leptoquark and encodes the Froggatt-Nielsen power counting, and $g_3$ is an overall coupling which is expected to be of $\mathcal{O}(1)$. Notice that the scalar leptoquark $S_1$ does not contribute to $b\to s\ell_1^-\ell_2^+$ transitions.
With this, we find 
\begin{align}
\label{eq:CsFN}
     C_9^{\mu\tau} = - C_{10}^{\mu\tau}=&\, -(0.41\pm0.07) \,, \nonumber\\
     C_9^{\tau\mu} = - C_{10}^{\tau\mu}=&\,\phantom{+}(10\pm2) \,.
\end{align}
For the modes with electrons and muons in the final states, we find $C_9^{e\mu}\propto 10^{-3}$ and a even lower value for $C_9^{\mu e}$. Therefore, we conclude that the corresponding branching ratios are too small to be measured by any experiment in the near future.\\
Focusing then on the final states with muons and taus, using the Wilson coefficients in \eqref{eq:CsFN} and our results in \sec{sec:2} gives
\begin{equation}
\begin{aligned}
    \mathcal{B}^{\mu\tau} =\,& (7.1\pm2.5 )\cdot 10^{-9}\,, \\
    \mathcal{B}^{\tau\mu} =\,& (4.2\pm 1.7)\cdot 10^{-6}\,,
\end{aligned}
\label{eq:BRFN}
\end{equation}
where the errors are dominated by the ones on the NP Wilson coefficients. Note that since this model predicts $C_9^{\ell_1\ell_2}=-C_{10}^{\ell_1\ell_2}$, $A_\mathrm{FB}^{\ell_1\ell_2}$ is independent from any Wilson coefficients and assumes the value in \Eq{eq:afbfix}. 
From \fig{fig:AFB} we can conclude that the $C_9^{\ell_1\ell_2}=-C_{10}^{\ell_1\ell_2}$ scenario would be excluded by the measurement of $A_\mathrm{FB}^{\ell_1\ell_2}$. Hence, this stresses the importance of obtaining experimental constraints on this observable. \\

\noindent {\bf The \boldmath$U_1$ vector leptoquark scenario {\cite{Cornella:2021sby}}}\\
Other interesting NP models are those with a vector leptoquark, usually denoted $U_1$. In fact, this NP particle is the only one able to accommodate both classes of $B$ anomalies on its own. Among the various possibilities available in the literature, we focus on \cite{Cornella:2021sby}, where the vector leptoquark is a massive state originating from the Spontaneus Symmetry Breaking of a gauge groupe larger than the SM one. As a consequence of the gauge representation of the $U_1$ vector leptoquark, not only vector and axial vector couplings, but also scalar and pseudoscalar couplings are generated. In particular, the latter are very useful in explaining the large discrepancies in $b\to c\tau\bar\nu$ data and as a consequence, generate sizeable $b\to s\ell_1\ell_2$ interactions. Therefore, we expect very different signatures for the $U_1$ model than the ones in the scalar leptoquark case. In Ref.~\cite{Cornella:2021sby} several cases are taken into account, where the flavour structure of the NP couplings has a $U(2)^5$ flavour symmetry \cite{Barbieri:2011ci} or not, and where (pseudo-)scalar couplings are present or not. In the following we report results for the case in which no $U(2)^5$ flavour symmetry is assumed. We note that using the scenario based on th $U(2)^5$ flavour symmetry yields very similar results. We also note that given the flavour structure assumed in Ref.~\cite{Cornella:2021sby}, the couplings of the vector leptoquark to electrons is zero, hence no effect is predicted for $\Lambda_b\to\Lambda e^{\pm}\mu^{\mp}$. In the notation of Ref.~\cite{Cornella:2021sby} we have
\begin{equation}
    \begin{aligned}
    C_{9}^{\ell_1\ell_2} = - C_{10}^{\ell_1\ell_2} =& +\frac{2\pi}{\alpha_\mathrm{em}|V_{tb}V^*_{ts}|}C_U\beta_L^{2\ell_1}(\beta_L^{3\ell_2})^*\,, \\
    C_{S}^{\ell_1\ell_2} = - C_{P}^{\ell_1\ell_2} =& +\frac{4\pi}{\alpha_\mathrm{em}|V_{tb}V^*_{ts}|}C_U\beta_L^{2\ell_1}(\beta_R^{3\ell_2})^*\,,
    \end{aligned}
\end{equation}
where $C_U$ is a normalisation constant which contains the mass of the vector leptoquark normalised to the electroweak vacuum-expectation value and the gauge coupling of the leptoquark.  The factor $\beta_{L(R)}^{j\beta}$ represents the coupling in flavour space to left(right)-handed fermions.  In the following we neglect the uncertainties on the fitted parameters obtained from \cite{Cornella:2021sby}  due to their large  and asymmetric distributions. Either way, this scenario provides a useful benchmark that allows us to predict the size of LFV $\Lambda_b\to\Lambda$ decays. We first look at the case $\beta_R^{3\beta}=0$. We find
\begin{equation}
    \begin{aligned}
    C_9^{\tau\mu} = -C_{10}^{\tau\mu} =&-5.93\,,\\ C_9^{\mu\tau} = -C_{10}^{\mu\tau} =&+2.90\,.
    \end{aligned}
\end{equation}
The predictions for $A_\mathrm{FB}^{\ell_1\ell_2}$ in this case are the same as in \Eq{eq:afbfix}. The corresponding integrated branching ratios are:
\begin{equation}
    \begin{aligned}
    \mathcal{B}^{\tau\mu} &=1.5 \times 10^{-6} 
    \\   \mathcal{B}^{\mu\tau} &= 3.6 \times 10^{-7} \,.
    \end{aligned}
\end{equation}
In the case where $\beta_R^{3\beta}\neq 0$, we find 
\begin{equation}
    \begin{aligned}
    C_9^{\tau\mu} = -C_{10}^{\tau\mu} =&-4.47\,, & C_S^{\tau\mu} = -C_{P}^{\tau\mu} =&\phantom{+}0\,, \\ C_9^{\mu\tau} = -C_{10}^{\mu\tau} =&\phantom{+}2.03\,, & C_S^{\mu\tau} = -C_{P}^{\mu\tau}=&\phantom{+} 4.06\,, 
    \end{aligned}
\end{equation}
which yields 
\begin{equation}
    \begin{aligned}
    \mathcal{B}^{\tau\mu} =&\phantom{+}8.5 \times 10^{-7}\quad \mathrm{and}\quad &  A_\mathrm{FB}^{\tau\mu} =&\phantom{+} 0.14\,, \\ 
    \mathcal{B}^{\mu\tau} =&\phantom{+}5.3 \times 10^{-7}\quad \mathrm{and} \quad &  A_\mathrm{FB}^{\mu\tau} =&\phantom{+} 0.12\,.
    \end{aligned}
\end{equation}
Some comments are in order. In the scenario where $C_{S(P)}^{\ell_1\ell_2}=0$, $A_\mathrm{FB}^{\ell_1\ell_2}$ is independent from the Wilson coefficients and its value is given in \Eq{eq:afbfix}. We find that $\mathcal{B}^{\tau\mu}>\mathcal{B}^{\mu\tau}$ due to a factor of two between the respective NP couplings. 
In the case where we have also $C_{S}^{\mu\tau}=-C_{P}^{\mu\tau}\neq0$, we find that $\mathcal {B}^{\mu\tau}$ is surprisingly small due to the negative interference between $C_{S}^{\mu\tau}$ and $C_{P}^{\mu\tau}$. On the other hand, $A_\mathrm{FB}^{\mu\tau}$ is found to be smaller than $A_\mathrm{FB}^{\tau\mu}$, hence providing a possible way to distinguish the different scenarios. 

\subsection{LHCb prospects}
The results found in the above Sections indicate that $\Lambda_b\to\Lambda \ell_1\ell_2$ decays are very good probes of physics beyond the SM and provide in certain scenarios complementary bounds with respect to the ones from $\bar{B}\to\ell_1\ell_2$ and $B^+\to K^+\ell_1\ell_2$ decays. Here we want to comment on the prospective for measurement of $\Lambda_b\to\Lambda \ell_1\ell_2$ decays at the LHCb experiment. \\
If we consider measurement carried out with the same dataset, we expect for the measured yields:
\begin{equation}
    \frac{\mathcal{N}(\Lambda_b\to\Lambda(\to p\pi)\ell_1\ell_2)}{\mathcal{N}(B^+\to K^+\ell_1\ell_2)} = \frac{\mathcal{B}(\Lambda_b\to\Lambda(\to p\pi)\ell_1\ell_2)|
    _\mathrm{theory}}{\mathcal{B}(B^+\to K^+\ell_1\ell_2)|_\mathrm{theory}} \frac{f_{\Lambda_b}}{f_{B^+}}r_{\Lambda_b/B^+}\,,
\end{equation}
where $f_{\Lambda_b}/f_{B^+}$ is the ratios of the fragmentation functions for the $\Lambda_b$ and the $B^+$ modes, respectively, and $r_{\Lambda_b/B^+}$ is a correction factor due to different reconstruction efficiencies. In Ref.~\cite{Aaij:2019pqz}, the ratio $f_{\Lambda_b}/{(f_u+f_d)}$ is measured. Using isospin relations, we can write $f_{\Lambda_b}/{f_{B^+}} = 2 f_{\Lambda_b}/{(f_u+f_d)} = (0.518 \pm 0.036)$.
The ratio of the predicted values of the theoretical branching ratios depends on the NP model and final state leptons. However, as we noted already in \sec{sec:3.1}, the branching ratios for the baryon and the meson case are very similar in size: therefore, for an order of magnitude estimate, we consider them to be equal. The last piece of information needed is the ratio $r_{\Lambda_b/B^+}$, that is difficult to estimate without a thorough simulation of the LHCb detector. However, in order to give an estimate, we use the information in Refs.~\cite{Aaij:2019wad,Aaij:2018gwm}, that are based on the same integrated luminosity. From these papers we extract 
\begin{equation}
    \frac{\mathcal{N}(\Lambda_b\to \Lambda(\to p \pi)\mu^-\mu^+)}{\mathcal{N}(B^+\to K^+\mu^-\mu^+) } \approx 0.31\,.
\end{equation}
This means that we expect the efficiency for the reconstruction of the $\Lambda_b$ to be roughly $1.67$ times less than that of the $B^+$, when also taking into account the fragmentation fractions effect. Hence we set $r_{\Lambda_b/B^+}=1.67$.  We expect that all the other correction factors due to the reconstruction of the leptons in the final state cancel out since we are comparing the same leptonic final states in both decays. This yields
\begin{equation}
    \mathcal{B}(\Lambda_b\to\Lambda(\to p\pi)\ell_1\ell_2)\approx 1.67 \frac{f_{\Lambda_b}}{f_{B^+}}\mathcal{B}(B^+\to K^+\ell_1\ell_2)\,.
\end{equation}
Using the current upper limit on $\mathcal{B}(B^+\to K^+\tau^+\mu^-)$, we thus expect that LHCb can reach the following sensitivity:
\begin{equation} \mathcal{B}(\Lambda_b\to\Lambda(\to p\pi)\mu^-\tau^+)\lesssim 6.5\cdot 10^{-5}\,,
\label{eq:LHCb_recast}
\end{equation}
for Run 1 and 2 datasets. In the above estimate, we have not included any correction for the trigger efficiency, which can be different for the baryonic and mesonic mode. The estimate in \Eq{eq:LHCb_recast} can be compared to the model dependent and model independent bounds on $\mathcal{B}(\Lambda_b\to\Lambda\tau^+\mu^-)$ found in the previous Sections. In particular, the expected upper bound from LHCb would already give better constraints than the corresponding ones from the mesonic decays, as illustrated in \fig{fig:ortho}. We also stress that future runs will improve the upper limit in \Eq{eq:LHCb_recast} of at least a factor of roughly two with Run 3 and a factor of three with further runs \cite{Bediaga:2018lhg}.

\section{Conclusions}
\label{sec:4}
In this paper, we present the first full analysis of $\Lambda_b\to \Lambda \ell_1^{-} \ell_2^{+}$ lepton flavour violating (LFV) decays in terms of possible new physics operators. The main results of this paper are \eqs{eq:a}{eq:c}, where the coefficients of the angular distributions for $\Lambda_b\to \Lambda \ell_1^{-} \ell_2^+$ decays are given. We study the interplay between the baryonic and mesonic searches for LFV, where for the latter upper limits are already available. We convert these upper limits into constraints on the branching ratio and forward-backward asymmetry for $\Lambda_b\to \Lambda\ell_1^{-} \ell_2^{+}$ decays. We find that the $\Lambda_b\to \Lambda\ell_1^{-} \ell_2^{+}$ decays provide different constraints on the new physics Wilson coefficients than $\bar{B}_s\to \ell_1^{-} \ell_2^{+}$ and $B^+\to K^+ \ell_1^{-} \ell_2^{+}$ decays, and have the potential to reduce the allowed parameter space for new physics models. We then analyse quantitatively the size of $\Lambda_b\to \Lambda\ell_1^{-} \ell_2^{+}$ decays in specific scenarios that can address $B$ anomalies, using as a reference \cite{Bordone:2020lnb} and \cite{Cornella:2021sby}. Our findings indicate that the predicted branching ratio for $\Lambda_b\to \Lambda\ell_1^{-} \ell_2^{+}$ for these scenarios are such that they can further constrain the new physics couplings. As a final prospective, we estimate the reach of LHCb for $\Lambda_b\to \Lambda\ell_1^{-} \ell_2^{+}$ decays, finding that an upper limit of $\mathcal{B}(\Lambda_b\to\Lambda\mu^-\tau^+)\lesssim 6.5\cdot 10^{-5}$ can be reached with Run 1 and Run 2 data.

\subsubsection*{Acknowledgements}
We thank Yasmine Amhis, Flavio Archilli, Lex Greeven and Mick Mulder for insightful discussion on experimental prospects. The work of MB is supported by the Italian Ministry of Research (MIUR) under grant PRIN 20172LNEEZ.  The work of MR is supported by the Deutsche Forschungsgemeinschaft (DFG, German Research Foundation) under grant  396021762 - TRR 257.

\appendix
\section{Details on kinematics}
\label{app:detkin}

In the $\Lambda_b$ rest frame ($\Lambda_b-\mathrm{RF}$), the momenta are defined as
\begin{align}
    q^\mu \vert_{\Lambda_b-\mathrm{RF}}=\,& (q^0,0,0,-|\vec{q}|)\,, \\
    k^\mu\vert_{\Lambda_b-\mathrm{RF}} =\,& (\mLamB-q^0,0,0,|\vec{q}|)\,.
\end{align}
where 
\begin{equation}
    q^0\vert_{\Lambda_b-\mathrm{RF}} = \frac{\mLamB^2-\mLam^2+q^2}{2\mLamB}\,, \quad \mathrm{and} \quad |\vec{q}|\vert_{\Lambda_b-\mathrm{RF}} = \frac{\sqrt{\lambda(\mLamB^2,\mLam^2,q^2)}}{2 \mLamB}\,,
\end{equation}
where $\lambda$ is the usual K\"allen function defined as $\lambda(a,b,c) = a^2+b^2+c^2-2 a (b+c)-2bc$.\\
In the dilepton rest frame  we have that $q^\mu\vert_{2\ell-\mathrm{RF}} = \sqrt{q^2}(1,0,0,0)$, and
\begin{align}
    p_1^\mu\vert_{2\ell-\mathrm{RF}} =\,& (E_{\ell_1},-|\vec{p}_2|\vert_{2\ell-\mathrm{RF}}\sin\theta_\ell,0,-|\vec{p}_2|\vert_{2\ell-\mathrm{RF}}\cos\theta_\ell)\,, \\
    p_2^\mu\vert_{2\ell-\mathrm{RF}} =\,& (E_{\ell_2},+|\vec{p}_2|\vert_{2\ell-\mathrm{RF}}\sin\theta_\ell,0,+|\vec{p}_2|\vert_{2\ell-\mathrm{RF}}\cos\theta_\ell)\,,
\end{align}
where
\begin{equation}
    |\vec{p}_2|\vert_{2\ell-\mathrm{RF}}   = \frac{\sqrt{\lambda(q^2,m_{\ell_1}^2,m_{\ell_2}^2)}}{2\sqrt{q^2}}\,, \quad \mathrm{and} \quad E_{\ell_{1,2}} = \frac{q^2+m_{\ell_{1,2}}^2-m_{\ell_{2,1}}^2}{2\sqrt{q^2}}.
\end{equation}
The two reference systems are connected by the following relation for any vector:
\begin{equation}
    x^\mu |\vert_{\Lambda_b-\mathrm{RF}} = \Lambda_{\mu\nu}  x^{T\mu} \, , \hspace{0.6cm} \Lambda = \begin{pmatrix}
\gamma & 0 & 0 & -\beta \gamma \\
0 & 1 & 0 & 0 \\
0 & 0 & 1 & 0 \\
-\beta \gamma & 0 & 0 & \gamma \\
\end{pmatrix}
\end{equation}
where $\Lambda_{\mu\nu}$ is a Lorentz transformation along the $z$ axis. 
It's parameters are:
\begin{equation}
    \gamma = \frac{q^0\vert_{\Lambda_b-\mathrm{RF}}}{\sqrt{q^2}}\,, \quad \mathrm{and} \quad \beta = \frac{|\vec{q}|\vert_{\Lambda_b-\mathrm{RF}}}{q^0\vert_{\Lambda_b-\mathrm{RF}}}
\end{equation}

\section{Correlations}
\label{sec:correlations}
We present correlation matrices for the set of coefficients $\{\xi_i^{\ell_1\ell_2},\rho_i^{\ell_1\ell_2}\}$, with the same ordering as in \tables{tab:resultsxi}{tab:resultsrho}. In \Table{tab:corrmue} we present the correlations for $\mu e$ final states and in \Table{tab:corrmutau} the ones for $\mu\tau$ final states.

\begin{table}[h]
\begin{center}
\centering
        \begin{tabular}{cccccccccc}
        \toprule
 1 & 1 & 0.617 & 0.617 & 0.643 & 0.643 &
   -0.728 & 0.820 & -0.839 & -0.839 \\
 1 & 1 & 0.617 & 0.617 & 0.643 & 0.643 &
   -0.728 & 0.820 & -0.839 & -0.839 \\
 0.617 & 0.617 & 1 & 1 & 0.885 & 0.885 &
   -0.559 & 0.451 & -0.778 & -0.778 \\
 0.617 & 0.617 & 1 & 1 & 0.885 & 0.885 &
   -0.559 & 0.451 & -0.778 & -0.778 \\
 0.643 & 0.643 & 0.885 & 0.885 & 1 & 1 &
   -0.835 & 0.438 & -0.911 & -0.911 \\
 0.643 & 0.643 & 0.885 & 0.885 & 1 & 1 &
   -0.835 & 0.437 & -0.911 & -0.911 \\
 -0.728 & -0.728 & -0.559 & -0.559 &
   -0.835 & -0.835 & 1 & -0.434 & 0.932 &
   0.932 \\
 0.820 & 0.820 & 0.451 & 0.451 & 0.438 &
   0.438 & -0.434 & 1 & -0.517 & -0.517 \\
 -0.839 & -0.839 & -0.778 & -0.778 & -0.911
   & -0.911 & 0.932 & -0.517 & 1 & 1 \\
 -0.839 & -0.839 & -0.778 & -0.778 &
   -0.911 & -0.911 & 0.932 & -0.517 & 1 &
   1 \\
   \toprule
   \end{tabular}
   \end{center}
\caption{Correlation matrix for the $\Lambda_b\to\Lambda \mu^-e ^+$ parameters.}
\label{tab:corrmue}
\end{table}

\begin{table}[t]
\begin{center}
        \begin{tabular}{cccccccccc}
        \toprule
 1 & 0.997 & 0.709 & 0.716 & -0.742 & 0.742 & 0.835 &
   0.857 & -0.877 & 0.877 \\
 0.997 & 1 & 0.747 & 0.755 & -0.787 & 0.788 & 0.858 &
   0.838 & -0.900 & 0.900 \\
 0.709 & 0.747 & 1.00 & 0.999 & -0.962 & 0.947 & 0.715 &
   0.466 & -0.835 & 0.835 \\
 0.716 & 0.755 & 0.999 & 1 & -0.971 & 0.958 & 0.734 &
   0.470 & -0.846 & 0.846 \\
 -0.742 & -0.787 & -0.962 & -0.971 & 1 & -0.999 &
   -0.841 & -0.481 & 0.899 & -0.899 \\
 0.742 & 0.788 & 0.947 & 0.958 & -0.999 & 1 & 0.857 &
   0.480 & -0.903 & 0.903 \\
 0.835 & 0.858 & 0.715 & 0.734 & -0.841 & 0.857 & 1 &
   0.519 & -0.964 & 0.964 \\
 0.857 & 0.838 & 0.466 & 0.470 & -0.481 & 0.480 & 0.519
   & 1 & -0.544 & 0.544 \\
 -0.877 & -0.900 & -0.835 & -0.846 & 0.899 & -0.903 &
   -0.964 & -0.544 & 1 & -1 \\
 0.877 & 0.900 & 0.835 & 0.846 & -0.899 & 0.903 & 0.964
   & 0.544 & -1 & 1 \\
 \toprule
   \end{tabular}
   \end{center}
\caption{Correlation matrix for the $\Lambda_b\to\Lambda \mu^-\tau ^+$ parameters.}
\label{tab:corrmutau}
\end{table}

\newpage
\bibliographystyle{JHEPtest} 
\bibliography{refs.bib}
\end{document}